\documentclass[twoside,11pt]{article}

\def\Rsp{{\mathbb{R}}}

\def\px{P_{\Xmatsc}}

\def\ps{P_{\Smatsc}}

\def\qx{Q^{\left(u\right)}_{\Xmatsc}}

\def\qxk{Q^{\left(u\right)}_{X_{k}}}

\def\qxEXP{Q^{\left(u_{\rm{E}}\right)}_{\Xmatsc}}

\def\uexp{u_{\expsc}}

\def\uGausss{u_{\Gausssc}}

\newcommand{\expsc}{{\mbox{\tiny${\rm{E}}$}}}

\newcommand{\Gausssc}{{\mbox{\tiny${\rm{G}}$}}}

\newcommand{\yvec}{{\bf{y}}}

\newcommand{\xvec}{{\bf{x}}}

\newcommand{\rvec}{{\bf{r}}}

\newcommand{\zerovec}{{\bf{0}}}

\newcommand{\Lambdamat}{{\bf{\Lambda}}}

\newcommand{\Amat}{{\bf{A}}}

\newcommand{\Bmat}{{\bf{B}}}

\newcommand{\Emat}{{\bf{E}}}

\newcommand{\Gmat}{{\bf{G}}}

\newcommand{\Hmat}{{\bf{H}}}

\newcommand{\Smat}{{\bf{S}}}

\newcommand{\Umat}{{\bf{U}}}

\newcommand{\Vmat}{{\bf{V}}}

\newcommand{\Wmat}{{\bf{W}}}

\newcommand{\Xmat}{{\bf{X}}}

\newcommand{\Xmatsc}{{\mbox{\boldmath \tiny $\Xmat$}}}

\newcommand{\Zmatsc}{{\mbox{\boldmath \tiny $\Zmat$}}}

\newcommand{\Smatsc}{{\mbox{\boldmath \tiny $\Smat$}}}

\newcommand{\Zmat}{{\bf{Z}}}

\def\bSigma{{\mbox{\boldmath $\Sigma$}}}

\def\psivec{{\mbox{\boldmath $\psi$}}}

\def\tvec{{\mbox{\boldmath $t$}}}

\def\XCalsc{{\mbox{\tiny $\mathcal{X}$}}}

\def\XCal{{\mbox{$\mathcal{X}$}}}

\def\SCal{{\mbox{$\mathcal{S}$}}}

\def\muvec{{\mbox{\boldmath $\mu$}}}

\newcommand{\be}{\begin{equation}}

\newcommand{\ee}{\end{equation}}

\newcommand{\beqna}{\begin{eqnarray}}

\newcommand{\eeqna}{\end{eqnarray}}

\usepackage{jmlr2e}
\usepackage{natbib}
\usepackage{graphicx}
\usepackage{amsmath,amssymb, amsthm, bbm}
\usepackage{epsfig}
\usepackage{subfigure}
\usepackage{arydshln}
\usepackage{paralist}
\usepackage{algpseudocode}
\usepackage{algorithm}

\newtheorem{Theorem}{Theorem}
\newtheorem{Lemma}{Lemma}

\newtheorem{Proposition}{Proposition}
\newtheorem{Definition}{Definition}

\firstpageno{1}

\begin{document}
\title{Measure Transformed Independent Component Analysis}

\author{\name Koby Todros \email ktodros@bgu.ac.il \\
       \addr Department of Electrical and Computer Engineering\\
       Ben-Gurion University of the Negev\\
       Beer-Sheva 84105, Israel
       \AND
       \name Alfred O. Hero \email hero@eecs.umich.edu \\
       \addr Department of Electrical Engineering and Computer Science\\
       University of Michigan\\
       Ann-Arbor 48105, MI, U.S.A}

\editor{}


\maketitle

\begin{abstract}
In this paper we derive a new framework for independent component analysis (ICA), called measure-transformed ICA (MTICA), that is based on applying a structured transform to the probability distribution of the observation vector, i.e., transformation of the probability measure defined on its observation space. By judicious choice of the transform we show that the separation matrix can be uniquely determined via diagonalization of several measure-transformed covariance matrices. In MTICA the separation matrix is estimated via approximate joint diagonalization of several empirical measure-transformed covariance matrices. Unlike kernel based ICA techniques where the transformation is applied repetitively to some affine mappings of the observation vector, in MTICA the transformation is applied only once to the probability distribution of the observations. This results in performance advantages and reduced implementation complexity. Simulations demonstrate the advantages of the proposed approach as compared to other existing state-of-the-art methods for ICA. 
\end{abstract}
\section{Introduction}
\label{Intro}
Independent component analysis (ICA) is a technique for multivariate data analysis that aims at decomposing an observed random vector into linear combination of mutually independent random variables \citep{Common, Oja}. The observation vector is assumed to be generated by an unknown linear mixture of mutually independent latent variables, called sources, with unknown distributions. The coefficient matrix of the linear mixture is called the mixing matrix and assumed to be invertible. Given a sequence of i.i.d. samples from the distribution of the observed vector, ICA aims to estimate the inverse of the mixing matrix, called the separation matrix, that is used for recovering the sources. Unlike principal component analysis, ICA can deal with a general mixing structure, which is not constrained to be orthogonal. The mutual independence assumption is plausible in a wide variety of fields, including telecommunications \citep{Communication,Communication2}, finance \citep{Finance, Finance2}, and biomedical signal analysis \citep{Bio, Bio2}, which makes ICA a natural tool for blind source separation in linear mixtures. 
 
ICA algorithms can be categorized as either parametric or semi-parametric. Parametric ICA methods involve specifying parametric models for the probability distributions of the sources followed by optimization of contrast functions that involve both the mixing matrix and the model's nuisance parameters. Generally, these contrast functions are  based on the likelihood function \citep{InfoMax, EInfoMax, Pham1, EFICA, Todros}, on non-Gaussianity measures such as kurtosis \citep{FastICA}, or on high-order correlations such as fourth-order cross-cumulants \citep{JADE, Cardoso1}. The main drawback of these techniques is that they might fail whenever the modeling assumptions are not satisfied. Unlike parametric ICA techniques, semi-parametric ICA methods \citep{Yeredor, Bach, RADICAL, NPICA, FKICA, EBMICA} assume nothing about the probability distributions of the sources, which make them more robust to varying source distributions.   

Another way to classify ICA algorithms is to divide them into data-based and statistically-based techniques. Data-based techniques \citep{InfoMax, FastICA, EInfoMax, Bach, RADICAL, NPICA, EFICA, FKICA, EBMICA} involve successive linear transformations that are applied to the data until some criterion of independence is maximized. These techniques require storage of the entire data record since it must be re-analyzed at each iteration. Unlike data-based techniques, in statistically-based methods \citep{ JADE, Common, SOBI, Yeredor, Todros}, the data is condensed into a smaller set of summary statistics that are computed only once. These summary statistics are then used to estimate the separation matrix.          
    
In this paper we introduce a new semi-parametric statistically-based ICA framework. The proposed framework, called measure-transformed ICA (MTICA), is inspired by a measure transformation approach that was recently applied to canonical correlation analysis \citep{MTCCA}. MTICA is based on applying a transform to the probability distribution of the observation vector, i.e., transformation of the probability measure defined on the observation space. The proposed transform is structured by a non-negative function called the MT-function. It preserves statistical independence and maps the probability distribution into a set of new probability measures on the observation space. By modifying the MT-function, classes of measure transformations can be obtained that have different useful properties. Under the proposed transform we define the measure-transformed (MT) covariance and derive its strongly consistent estimate, which is also shown to be Fisher consistent \citep{Cox}. Robustness of the empirical MT-covariance to outliers is studied by analyzing its influence function \citep{Hampel}. A sufficient condition on the MT-function that guarantees B-robustness of the empirical-MT covariance is established. In MTICA the separation matrix is estimated via approximate joint diagonalization \citep{FG, EJM, Pham, Yeredor2, FFDIAG, QDIAG, Fadaili, FAJD, FastJDWM, TodrosJD} of several empirical measure-transformed covariance matrices.

The MT-function are selected from either exponential or Gaussian families of functions parameterized by scale and/or translation parameters. When we use an exponential MT-function the corresponding measure-transformed covariance matrix of the observation vector is equal to the Hessian of the cumulant-generating-function, resulting in the ICA method proposed in \citep{Yeredor}, which we call here exponential-MTICA. In \citep{Yeredor} the author showed that if at most one of the sources is Gaussian, then the mixing matrix can be uniquely identified, up to scaling and permutations of its columns, via non-symmetric eigenvalue decomposition that involves two Hessians of the cumulant-generating-function. Based on this property, exponential-MTICA estimates the separation matrix via non-orthogonal approximate joint diagonalization (NOAJD) 
\citep{Pham, Yeredor2, FFDIAG, QDIAG, Fadaili, FAJD, TodrosJD} over a set of empirical exponential MT-covariance matrices. These matrices are obtained by evaluating the exponential MT-function at different test-points in the parameter space.  
   
When we use a Gaussian MT-function a new algorithm for ICA, called Gaussian-MTICA, is obtained. We show that if at most one of the sources is Gaussian, then 
the unitary mixing matrix associated with the whitened observation vector can be uniquely identified via symmetric eigenvalue decomposition of a single Gaussian MT-covariance matrix. Gaussian-MTICA estimates the separation matrix via empirical whitening and orthogonal approximate joint diagonalization (OAJD) \citep{FG, EJM} over a set of empirical Gaussian MT-covariance matrices. As in exponential-MTICA, these matrices are obtained by evaluating the Gaussian MT-function at different test-points in the parameter space. 

In the paper we show that identifiability of the mixing matrices in the exponential-MTICA and Gaussian-MTICA algorithms is based on the following measure-transformation invariance properties: 
\begin{inparaenum}[\upshape(1\upshape)]
\item
The Gaussian family of distributions is closed under measure transformations generated by the exponential or Gaussian MT-functions.
\item
A random variable is Gaussian if and only if its measure-transformed variance generated by the exponential or Gaussian MT-functions is constant over any open interval defined over their scaling and translation parameter axes.
\end{inparaenum}

MTICA has the following advantages over existing state-of-the-art ICA methods:
\begin{inparaenum}[\upshape(1\upshape)]   
\item
Similarly to other semi-parametric ICA techniques, such as kernel-ICA-KGV (KGV) \citep{Bach} and RADICAL \citep{RADICAL}, MTICA do not rely on restrictive assumptions about the distribution of the sources. Therefore, unlike parametric ICA methods such as fast-ICA (FICA) \citep{FastICA}, efficient fast-ICA (EFICA) \citep{EFICA}, JADE \citep{JADE} and extended Infomax (EIMAX) \citep{EInfoMax}, the MTICA methods are more robust to varying source distributions.
\item
MTICA is comprised of a non-iterative part for estimation of the MT-covariance matrices followed by an iterative part for performing approximate joint diagonalization. The non-iterative part has computational complexity that is linear in the sample size while the computational complexity of the iterative part is sample size independent. This results in reduced computational complexity in comparison to data-based techniques such as KGV and RADICAL whose computational complexity is super-linear in the sample size.
\item
In contrast to KGV the MTICA techniques do not expand the dimension of the observed vector, nor do they require regularization of the measure-transformed covariance matrices. 
\item
Unlike KGV that involves complex optimization over the Stiefel manifold \citep{Edelman}, the MTICA methods are easy to implement and only involve simple estimation of some MT-covariance matrices followed by approximate joint diagonalization which can be performed with off-the-shelf algorithms \citep{FG, EJM, Pham, Yeredor2, FFDIAG, QDIAG, Fadaili, FAJD, FastJDWM, TodrosJD}. 
\item
The Gaussian MT-function is bounded and has the property that it de-emphasizes samples distant from its location parameter. Consequently, unlike cumulant based techniques such as JADE, FICA, and EFICA the Gaussian-MTICA is highly robust to outliers. This property is supported by the fact that the empirical Gaussian MT-covariance, whose influence function is bounded, is B-robust. 
\item
Unlike ICA techniques that are based on whitening and unitary de-mixing, the exponential-MTICA algorithm is more robust to model mismatch scenarios where the whitened observations do not admit unitary mixing. 
\end{inparaenum}

The proposed MTICA approach is evaluated by simulation to illustrate its advantages relative to other state-of-the-art ICA techniques, such as 
FICA, EFICA, JADE, EIMAX, KGV, and RADICAL.

The paper is organized as follows. In Section \ref{PRE}, we review the ICA problem. In Section \ref{MTICA}, the MTICA procedure is derived. In Section \ref{EMTICA}, the exponential-MTICA method and its relation to \citep{Yeredor} are discussed. In Section \ref{GMTICA}, the Gaussian-MTICA method is developed. Comparisons between exponential-MTICA and Gaussian-MTICA are given in Section \ref{EMTICAvsGMTICA}. In Section \ref{ACL}, the computational complexity of the MTICA algorithms is discussed and compared to those of other ICA techniques. In Section \ref{Examp}, the performance of the proposed approach is compared to other ICA techniques via simulation experiments. In Section \ref{Disc}, the main points of this contribution are summarized. The propositions and theorems stated throughout the paper are proved in the Appendices.
\section{Independent component analysis: Review}
\label{PRE}    
\subsection{Preliminaries}
Let $\Xmat=\left[X_{1},\ldots,X_{p}\right]^{T}$ denote a random vector, whose observation space is $\XCal\subseteq\Rsp^{p}$. We define the measure space $\left(\XCal,\mathcal{S}_{\XCalsc},\px\right)$, where $\mathcal{S}_{\XCalsc}$ is a $\sigma$-algebra over $\XCal$, and $\px$ is the joint probability measure on $\mathcal{S}_{\XCalsc}$. Let $\XCal_{k}$ denote the observation space of $X_{k}$. The marginal probability measure of $\px$ on $\mathcal{S}_{\XCalsc_{k}}$ is denoted by $P_{X_{k}}$, were  $\mathcal{S}_{\XCalsc_{k}}$ is the $\sigma$-algebra over $\XCal_{k}$. Let $g\left(\cdot\right)$ denote an integrable scalar function on $\XCal$. The expectation of $g\left(\Xmat\right)$ under $\px$ is defined as
\begin{equation}
\label{ExpDef}
{\rm{E}}\left[g\left(\Xmat\right);\px\right]\triangleq\int\limits_{\XCalsc}g\left(\xvec\right)d\px\left(\xvec\right),
\end{equation}
where $\xvec\in\XCal$. The components of $\Xmat$ are mutually independent under $\px$ if
\begin{equation}
\label{MutInd}
{\rm{E}}\left[g\left(X_{j}\right)h\left(X_{k}\right);\px\right]={\rm{E}}\left[g\left(X_{j}\right);P_{X_{j}}\right]{\rm{E}}\left[h\left(X_{k}\right);P_{X_{k}}\right]\hspace{0.2cm}\forall{j\neq{k}},
\end{equation}
for all integrable scalar functions $g\left(\cdot\right)$, $h\left(\cdot\right)$ on $\XCal$. The components of $\Xmat$ are mutually uncorrelated under $\px$ if 
\begin{equation}
\label{MutCorr}
{\rm{E}}\left[X_{j}X_{k};\px\right]={\rm{E}}\left[X_{j};P_{X_{j}}\right]{\rm{E}}\left[X_{k};P_{X_{k}}\right]\hspace{0.2cm}\forall{j\neq{k}}.
\end{equation}
The empirical distribution, $\hat{\px}$, based on a sequence of samples $\Xmat_{n}$, $n=1,\ldots,N$, is specified by
\begin{equation}
\label{EmpProbMes}
\hat{P}_{\Xmatsc}\left(A\right)=\frac{1}{N}\sum\limits_{n=1}^{N}\delta_{\Xmatsc_{n}}\left(A\right),
\end{equation}
where $A\in\mathcal{S}_{\XCalsc}$ and $\delta_{\Xmatsc_{n}}\left(\cdot\right)$ is the Dirac probability measure at $\Xmat_{n}$ \citep{Folland}.
\subsection{Independent component analysis}
The instantaneous noiseless ICA model takes the following form:  
\begin{equation}
\label{ICAModel}
\Xmat=\Amat\Smat,
\end{equation}
where $\Xmat\in\Rsp^{p}$, $p\geq{2}$, is an observed random vector, $\Amat\in\Rsp^{{p}\times{p}}$ is an invertible unknown matrix, called the mixing matrix, and $\Smat\in\Rsp^{p}$ is a latent random vector comprised of mutually independent variables having finite second-order moments and unknown distributions. The components of $\Smat$ are also called sources. Under the model (\ref{ICAModel}) it has been shown that the mixing matrix $\Amat$ can be uniquely identified, up to permutation and scaling of its columns, if and only if at most one of the sources is Gaussian \citep{Kagan, Common, Oja, Visa}. Given a sequence of i.i.d. samples from $\px$, ICA aims to estimate the separation matrix $\Bmat=\Amat^{-1}$ and thus, recover the sources using the relation $\Smat=\Bmat\Xmat$.

Many state-of-the-art ICA algorithms, such as JADE, FICA, EFICA, EIMAX, KGV and RADICAL, referenced in Section \ref{Intro}, apply whitening to the observed vector $\Xmat$. The whitened observation vector is represented as
\begin{equation}
\label{ZDef}
\Zmat\triangleq\Wmat\Xmat=\Umat\Smat,
\end{equation}
where $\Wmat\in\Rsp^{p\times{p}}$ is the whitening matrix and $\Umat\triangleq\Wmat\Amat$. Assuming, without loss of generality, that the components of $\Smat$ have unit variances, one can easily verify that the matrix $\Umat$ is unitary leading to a unitary mixing model. Let $\Vmat\triangleq\Umat^{T}$, where $(\cdot)^{T}$ denotes the transpose operator. ICA algorithms that use whitening implement an estimate of $\Vmat$ using constraint optimization over the Stiefel manifold of unitary matrices \citep{Edelman}. The empirical separation matrix is then obtained using the relation $\Bmat=\Vmat\Wmat$.
\section{Measure transformed ICA}
\label{MTICA}
In this section the MTICA procedure is presented. First, a transform that maps a probability measure $\px$ into a set of probability measures $\left\{\qx\right\}$ on $\mathcal{S}_{\XCalsc}$ is defined that has the property that it preserves mutual independence between the components of $\Xmat$ under $\px$. Second, we define the measure-transformed covariance and derive its strongly consistent estimate, which is also shown to be Fisher consistent \citep{Cox}. Robustness of the empirical measure-transformed covariance to outliers is studied by analyzing the boundedness of its influence function \citep{Hampel}. Finally, based on the mixing models (\ref{ICAModel}), (\ref{ZDef}), the MTICA procedure is specified by performing approximate joint diagonalization of a set of empirical measure-transformed covariance matrices.
\subsection{Probability measure transform}
\begin{Definition}
\label{Def1}
Given a probability measure $\px$ and a non-negative function $u:\Rsp^{p}\rightarrow\Rsp_{+}$ satisfying  
\begin{equation}
\label{Assumption1}  
u\left(\Xmat\right)=\prod\limits_{k=1}^{p}{u_{k}}\left(X_{k}\right),\hspace{0.4cm}u_{k}:\Rsp\rightarrow\Rsp_{+},\hspace{0.2cm}k=1,\ldots,p,
\end{equation}
and
\begin{equation}
\label{Assumption2}  
0<{{{\rm{E}}}\left[u\left(\Xmat\right);\px\right]}<\infty,
\end{equation}
a transform on $\px$ is defined via the following relation:
\begin{equation}
\label{MeasureTransform} 
\qx\left(A\right)\triangleq{\rm{T}}_{u}\left[\px\right]\left(A\right)=\int\limits_{A}\varphi_{u}\left(\xvec\right)d\px\left(\xvec\right),
\end{equation}
where $A\in\mathcal{S}_{\XCalsc}$, $\xvec=\left[x_{1},\ldots,x_{p}\right]^{T}\in\XCal$, and
\begin{equation}
\label{VarPhiDef} 
\varphi_{u}\left(\xvec\right)\triangleq\frac{u\left(\Xmat\right)}{{{\rm{E}}}\left[u\left(\Xmat\right);\px\right]}.
\end{equation}
The function $u\left(\cdot\right)$, associated with the transform ${\rm{T}}_{u}\left[\cdot\right]$, is called the MT-function.
\end{Definition}  
In the following Proposition, some properties of the measure transform (\ref{MeasureTransform}) are given.   
\begin{Proposition}
\label{Prop1}
Let $\qx$ be defined by relation (\ref{MeasureTransform}). 
Then
\begin{enumerate}[\upshape(1\upshape)]
\item
\label{P1}
$\qx$ is a probability measure on $\mathcal{S}_{\XCalsc}$.
\item
\label{P2}
$\qx$ is absolutely continuous w.r.t. $\px$, with Radon-Nikodym derivative \citep{Folland} given by
\begin{equation}
\label{MeasureTransformRadNik}   
\frac{d\qx\left(\xvec\right)}{d\px\left(\xvec\right)}=\varphi_{u}\left(\xvec\right).
\end{equation}
\item
\label{P21}
Assume that the MT-function $u\left(\cdot\right)$ is strictly positive, then $\px$ is absolutely continuous w.r.t. $\qx$ with a strictly positive Radon-Nikodym derivative given by 
\begin{equation}
\label{RadNik2} 
\frac{d\px\left(\xvec\right)}{d\qx\left(\xvec\right)}={\varphi^{-1}_{u}\left(\xvec\right)}=\frac{u^{-1}\left(\xvec\right)}{{\rm{E}}\left[u^{-1}\left(\Xmat\right);\qx\right]}.
\end{equation}
\item 
\label{P3} 
If $X_{1},\ldots,X_{p}$ are mutually independent under $\px$, then they are mutually independent under $\qx$.
\end{enumerate} 
[A proof is given in Appendix \ref{Prop1Proof}]
\end{Proposition}
The probability measure $\qx$ is said to be generated by the MT-function $u\left(\cdot\right)$. By modifying $u\left(\cdot\right)$, such that the conditions (\ref{Assumption1}), (\ref{Assumption2}) are satisfied, virtually any probability measure on $\mathcal{S}_{\XCalsc}$ can be obtained. 
\subsection{The measure-transformed covariance}
According to (\ref{ExpDef}) and (\ref{MeasureTransformRadNik}) the measure-transformed covariance of $\Xmat$ under $\qx$ is given by
\begin{equation} 
\label{MTCovZ}  
\bSigma^{\left(u\right)}_{\Xmatsc}={\rm{E}}\left[\Xmat\Xmat^{T}\varphi_{u}\left(\Xmat\right);\px\right]-\muvec^{\left(u\right)}_{\Xmatsc}\muvec^{\left(u\right)T}_{\Xmatsc},
\end{equation}
where
\begin{equation} 
\label{MTMeanZ}   
\muvec^{\left(u\right)}_{\Xmatsc}\triangleq{\rm{E}}\left[\Xmat\varphi_{u}\left(\Xmat\right);\px\right]
\end{equation}
is the measure-transformed expectation of $\Xmat$ under $\qx$. Equation (\ref{MTCovZ}) implies that $\bSigma^{\left(u\right)}_{\Xmatsc}$ is a weighted covariance matrix of $\Xmat$ under $\px$, with weighting function $\varphi_{u}\left(\cdot\right)$. Hence, $\bSigma^{\left(u\right)}_{\Xmatsc}$ can be estimated using only samples from the distribution $\px$. By modifying the MT-function $u\left(\cdot\right)$, such that the conditions (\ref{Assumption1}), (\ref{Assumption2}) are satisfied,  the MT-covariance matrix under $\qx$ is modified. In particular, by choosing $u\left(\xvec\right)\equiv{1}$, we have $\qx=\px$, and the standard covariance matrix is obtained. 

In the following Proposition a strongly consistent estimate of the measure-transformed covariance is given that is based on i.i.d. samples from the probability distribution $\px$.
\begin{Proposition}
\label{ConsistentEst}
Let $\Xmat_{n}$, $n=1,\ldots,N$ denote a sequence of i.i.d. samples from the distribution $\px$, and define the empirical covariance estimate
\begin{equation}
\label{Rx_u_Est}
\hat{\bSigma}^{\left(u\right)}_{\Xmatsc}\triangleq\sum\limits_{n=1}^{N}\Xmat_{n}\Xmat^{T}_{n}\hat{\varphi}_{u}\left(\Xmat_{n}\right)
-\hat{\muvec}^{\left(u\right)}_{\xvec}\hat{\muvec}^{\left(u\right)T}_{\xvec},
\end{equation}
where
\begin{equation}
\label{Mu_x_Mu_y}
\hat{\muvec}^{\left(u\right)}_{\Xmatsc}\triangleq\sum\limits_{n=1}^{N}\Xmat_{n}\hat{\varphi}_{u}\left(\Xmat_{n}\right),
\end{equation}
and
\begin{equation}
\label{varphi_hat}
\hat{\varphi}_{u}\left(\Xmat_{n}\right)\triangleq\frac{u\left(\Xmat_{n}\right)}{\sum\limits_{n=1}^{N}u\left(\Xmat_{n}\right)}.
\end{equation}
Assume 
\begin{eqnarray}
\label{Cond1}
{\rm{E}}\left[u^{2}\left(\Xmat\right);\px\right]<\infty
&{\rm{and}}& 
{\rm{E}}\left[X^{4}_{k};\px\right]<\infty\hspace{0.2cm}\forall{k=1,\ldots,p}.
\end{eqnarray}
Then $\hat{\bSigma}^{\left(u\right)}_{\Xmatsc}\rightarrow{\bSigma}^{\left(u\right)}_{\Xmatsc}$ almost surely as $N\rightarrow\infty$. [The proof is similar to the
proof of Proposition 3 in \citep{MTCCA} and therefore is omitted]
\end{Proposition}
Note that for $u\left(\Xmat\right)\equiv{1}$ the estimator $\frac{N}{N-1}\hat{\bSigma}^{\left(u\right)}_{\Xmatsc}$ reduces to the standard unbiased estimator of the covariance matrix $\bSigma_{\Xmatsc}$. Also notice that $\hat{\bSigma}^{\left(u\right)}_{\Xmatsc}$ can be written as a statistical functional $\textrm{H}_{u}\left[\cdot\right]$ of the empirical probability measure $\hat{P}_{\Xmatsc}$ (\ref{EmpProbMes}), i.e.,
\begin{equation}
\label{StatFunc}
\hat{\bSigma}^{\left(u\right)}_{\Xmatsc}=\frac{{\rm{E}}\left[\Xmat\Xmat^{T}{u}\left(\Xmat\right);\hat{P}_{\Xmatsc}\right]}{{\rm{E}}\left[{u}\left(\Xmat\right);\hat{P}_{\Xmatsc}\right]}-\frac{{\rm{E}}\left[\Xmat{u}\left(\Xmat\right);\hat{P}_{\Xmatsc}\right]{\rm{E}}\left[\Xmat^{T}{u}\left(\Xmat\right);\hat{P}_{\Xmatsc}\right]}{{\rm{E}^{}}^{2}\left[{u}\left(\Xmat\right);\hat{P}_{\Xmatsc}\right]}\triangleq{\textrm{H}}_{u}\left[\hat{P}_{\Xmatsc}\right].
\end{equation} 
According to (\ref{VarPhiDef}), (\ref{MTCovZ}), (\ref{MTMeanZ}), and (\ref{StatFunc}), when $\hat{P}_{\Xmatsc}$ is replaced by the true probability measure ${P}_{\Xmatsc}$ we have $\textrm{H}_{u}\left[P_{\Xmatsc}\right]=\bSigma^{\left(u\right)}_{\Xmatsc}$, which implies that $\hat{\bSigma}^{\left(u\right)}_{\Xmatsc}$ is Fisher consistent \citep{Cox}.
\subsection{Robustness of the empirical MT-covariance to outliers}
Here, we study the robustness of the empirical MT-covariance $\hat{\bSigma}^{\left(u\right)}_{\Xmatsc}$ to outliers using its influence function.
Define the probability measure $P_{\epsilon}\triangleq\left(1-\epsilon\right)\px+\epsilon\delta_{\yvec}$, where $0\leq\epsilon\leq1$, $\yvec\in\Rsp^{p}$, and
$\delta_{\yvec}$ is a Dirac probability measure at $\yvec$. The influence function of a Fisher consistent estimator with statistical functional $\textrm{H}\left[\cdot\right]$ at probability distribution $\px$ is defined pointwise as \citep{Hampel}:
\begin{equation}
\label{IFDef}
{\rm{IF}}_{\textrm{H},\px}\left(\yvec\right)\triangleq\lim\limits_{\epsilon\rightarrow{0}}\frac{\textrm{H}\left[P_{\epsilon}\right]-\textrm{H}\left[\px\right]}{\epsilon}
=\left.\frac{\partial{\textrm{H}}\left[{P}_{\epsilon}\right]}{\partial{\epsilon}}\right|_{\epsilon=0}.
\end{equation}
The influence function describes the effect on the estimator of an infinitesimal contamination at the point $\yvec$. An estimator is said to be B-robust if its influence function is bounded. Using (\ref{VarPhiDef}), (\ref{MTCovZ}), (\ref{MTMeanZ}), (\ref{StatFunc}) and (\ref{IFDef}) one can verify that the influence function of the empirical MT-covariance is given by
\begin{equation}
\label{MT_COV_INF}
{\rm{IF}}_{\textrm{H}_{u},\px}\left(\yvec\right)
=\frac{u\left(\yvec\right)}{{\rm{E}}\left[u\left(\Xmat\right);\px\right]}
\left( \left(\yvec-\muvec^{\left(u\right)}_{\Xmatsc}\right)\left(\yvec-\muvec^{\left(u\right)}_{\Xmatsc}\right)^{T} -  {\bSigma}^{\left(u\right)}_{\Xmatsc}  \right).
\end{equation}
The following proposition states a sufficient condition for boundedness of (\ref{MT_COV_INF}).
\begin{Proposition}
\label{RobustnessConditions}
The influence function (\ref{MT_COV_INF}) is bounded if the MT-function $u(\yvec)$ is bounded, and there exists a constant $c>0$ such that $u(\yvec)\leq{c}\|\yvec\|^{-2}_{2}$, where $\|\cdot\|_{2}$ denotes the $l_{2}$-norm. 
[A proof is given in Appendix \ref{RobustnessConditionsProof}]
\end{Proposition}
In Section \ref{GMTICA} we show that MT-functions chosen from the Gaussian family of functions satisfy these conditions, resulting in a measure-transformed ICA algorithm that is resilient to outliers.
\subsection{The MTICA procedure}
In MTICA we choose a sequence of MT-functions $u_{m}\left(\cdot\right)$, $m=1,\ldots,M$ that satisfies at least one of the following conditions: 
\begin{enumerate}[\upshape(1\upshape)]
\item
Under the ICA model (\ref{ICAModel}) the separation matrix $\Bmat$ is the unique matrix (up to permutation and scaling of its rows) that jointly diagonalizes the  MT-covariance matrices $\bSigma^{\left(u_{m}\right)}_{\Xmatsc}$, $m=1,\ldots,M$.
\item
Under the unitary mixing model (\ref{ZDef}) the 
matrix $\Vmat=\Umat^{T}$ is the unique matrix (up to permutation and sign of its rows) that jointly diagonalizes the MT-covariance matrices $\bSigma^{\left(u_{m}\right)}_{\Zmatsc}$, $m=1,\ldots,M$.
\end{enumerate}

When the first condition is satisfied, the separation matrix $\Bmat$ is estimated via NOAJD of the empirical MT-covariances $\hat{\bSigma}^{\left(u_{m}\right)}_{\Xmatsc}$, $m=1,\ldots,M$. The NOAJD \citep{Pham, Yeredor2, FFDIAG, QDIAG, Fadaili, FAJD, TodrosJD} seeks a non-singular matrix $\hat{\Bmat}\in\Rsp^{p\times{p}}$, such that  $\hat{\Bmat}\hat{\bSigma}^{\left(u_{m}\right)}_{\Xmatsc}\hat{\Bmat}^{T}$, $m=1,\ldots,M$ are ``as diagonal as possible'' in the sense that a deviation measure from diagonality is minimized. 
 The MTICA procedure in this case is summarized in Algorithm \ref{EMTICAAlg}.
\begin{algorithm}[htbp!]
\caption{MTICA with no whitening}\label{EMTICAAlg}
\textbf{Input:} A sequence of data samples $\Xmat_{n}$, $n=1,\ldots,N$.
\begin{algorithmic}[1]  
\State \label{S1EMTICA} Choose a sequence of MT-functions $u_{m}\left(\cdot\right)$, $m=1,\ldots,M$, such that $\Bmat$ is the unique joint diagonalization marix of $\bSigma^{\left(u_{m}\right)}_{\Xmatsc}$, $m=1,\ldots,M$.
\State Using (\ref{Rx_u_Est})-(\ref{varphi_hat})  derive the empirical MT-covariances $\hat{\bSigma}^{\left(u_{m}\right)}_{{\Xmatsc}}$, $m=1,\ldots,M$.
\State Find the NOAJD matrix $\hat{\Bmat}$ of $\hat{\bSigma}^{\left(u_{m}\right)}_{\Xmatsc}$, $m=1,\ldots,M$.
\end{algorithmic}
\textbf{Output:}  The empirical separation matrix $\hat{\Bmat}$.
\end{algorithm}

Alternatively, when the second condition is satisfied the observations are whitened, and the estimate of $\Vmat$ is obtained via OAJD of the empirical MT-covariance matrices $\hat{\bSigma}^{\left(u_{m}\right)}_{\hat{\Zmatsc}}$, $m=1,\ldots,M$, where $\hat{\Zmat}\triangleq\hat{\Wmat}\Xmat$ and $\hat{\Wmat}$ is the empirical whitening matrix. The OAJD \citep{FG, EJM} seeks a unitary matrix $\hat{\Vmat}\in\Rsp^{p\times{p}}$, such that  $\hat{\Vmat}\hat{\bSigma}^{\left(u_{m}\right)}_{\hat{\Zmatsc}}\hat{\Vmat}^{T}$, $m=1,\ldots,M$ are ``as diagonal as possible'' by, once again, minimizing a deviation measure from diagonality. The empirical separation matrix is obtained by taking $\hat{\Bmat}=\hat{\Vmat}\hat{\Wmat}$. The MTICA procedure in this case is summarized in Algorithm \ref{GMTICAAlg}.
\begin{algorithm}
\caption{MTICA with whitening}\label{GMTICAAlg}  
\textbf{Input:} A sequence of data samples $\Xmat_{n}$, $n=1,\ldots,N$.
\begin{algorithmic}[1]
\State \label{S1GMTICA} Choose a sequence of MT-functions $u_{m}\left(\cdot\right)$, $m=1,\ldots,M$, such that $\Vmat$ is the unique joint diagonalization matrix of $\bSigma^{\left(u_{m}\right)}_{\Zmatsc}$, $m=1,\ldots,M$. 
\State Estimate the whitening matrix $\hat{\Wmat}$.
\State Generate the sequence $\hat{\Zmat}_{n}=\hat{\Wmat}\Xmat_{n}$, $n=1,\ldots,N$.
\State Using (\ref{Rx_u_Est})-(\ref{varphi_hat}) derive the empirical MT-covariances $\hat{\bSigma}^{\left(u_{m}\right)}_{\hat{\Zmatsc}}$, $m=1,\ldots,M$.
\State Find the OAJD matrix $\hat{\Vmat}$ of $\hat{\bSigma}^{\left(u_{m}\right)}_{\hat{\Zmatsc}}$, $m=1,\ldots,M$.
\end{algorithmic}
\textbf{Output:}  Obtain an estimate of $\Bmat$ by taking $\hat{\Bmat}=\hat{\Vmat}\hat{\Wmat}$.
\end{algorithm}

By modifying the MT-functions such that the stated conditions are satisfied a family of measure-transformed independent component analyses can be obtained. Particular choices of MT-functions leading to the exponential and Gaussian MTICA algorithms are discussed in the next sections.
\section{Exponential-MTICA}
\label{EMTICA}
In this section we parameterize the MT-function $u\left(\cdot;\tvec\right)$, with scaling parameter $\tvec\in\Rsp^{p}$ under the exponential family of functions. 
Under this choice of MT-function the MT-covariance is given by the Hessian of the cumulant-generating-function resulting in the ICA algorithm proposed in \citep{Yeredor}. 
\subsection{The exponential MT-covariance matrix}
\label{ExpMTCov}
Let $u_{\rm{E}}\left(\cdot;\cdot\right)$ be defined as the parameterized function
\begin{equation}
\label{ExpMTFunc} 
\uexp\left(\xvec;\tvec\right)\triangleq\exp\left(\tvec^{T}\xvec\right),
\end{equation}
where $\tvec\in\Rsp$. Using (\ref{VarPhiDef}), (\ref{MTCovZ}) and (\ref{ExpMTFunc}) one can easily verify that the covariance matrix of $\Xmat$ under $\qxEXP$ takes the form
\begin{equation}
\label{OffHessz}
\bSigma^{\left(\uexp\right)}_{\Xmatsc}\left(\tvec\right)=\frac{\partial^{2}\log{M}_{\Xmatsc}\left(\tvec\right)}{\partial\tvec\partial\tvec^{T}},
\end{equation}
where
\begin{equation}
\label{MomGenFunc}  
{M}_{\Xmatsc}\left(\tvec\right)\triangleq{\rm{E}}\left[\exp\left(\tvec^{T}\Xmat\right);\px\right]
\end{equation}
is the  moment generating function of $\Xmat$, and it is assumed that ${M}_{\Xmatsc}\left(\tvec\right)$ is finite in some open region in $\Rsp^{p}$ containing the origin. Note that the covariance matrix in (\ref{OffHessz}) involves higher-order statistics of $\Xmat$. Additionally, observe that $\bSigma^{\left(\uexp\right)}_{\Xmatsc}\left(\tvec\right)$ reduces to the standard covariance matrix $\bSigma_{\Xmatsc}$ for $\tvec=\zerovec$. 

In the following lemma, directly following from (\ref{VarPhiDef})-(\ref{RadNik2}) and the definition of the exponential MT-function (\ref{ExpMTFunc}), one sees that the Gaussian family of probability measures is closed under the measure transformation (\ref{MeasureTransform}) when generated by the exponential MT-function. 
\begin{Lemma}
\label{ExpPresProp}
A random vector $\Xmat$ is Gaussian under the probability measure $\px$ if and only if it remains Gaussian under the transformed probability measure $Q^{\left(\uexp\right)}_{\Xmatsc}$ generated by the exponential MT-function $\uexp\left(\cdot;\cdot\right)$.
\end{Lemma}

This property is used in proving the following theorem that states a necessary and sufficient condition for Gaussianity of a random variable $X$ based on its exponential MT-variance.
\begin{Theorem}
\label{ConstVarExpMT}
A random variable $X$ with corresponding probability measure $P_{X}$ is Gaussian if and only if the the exponential MT-variance satisfies 
\begin{equation}
\label{ExpMTVarConst}
\sigma^{\left(\uexp\right)}_{X}\left(t\right)=c\hspace{0.2cm}\forall{t}\in\left(t_{0}-\epsilon,t_{0}+\epsilon\right),
\end{equation}
where $c$ and $\epsilon$ are some positive constants and $t_{0}$ is an arbitrary point in $\Rsp$. [A proof is given in Appendix \ref{ExpMTVarConstProof}]
\end{Theorem}
Hence, if a random variable $X$ is non-Gaussian then its exponential MT-variance $\sigma^{\left(\uexp\right)}_{X}\left(t\right)$ cannot be constant over any open interval. This property is used in the following subsection to establish identifiability of the mixing matrix $\Amat$. 
\subsection{Identifiability of the mixing matrix $\Amat$ under two exponential MT-covariance matrices}
Using (\ref{ICAModel}), (\ref{VarPhiDef}), (\ref{MTCovZ}) and (\ref{ExpMTFunc}) it can be shown that for any choice of the scaling parameter $\tvec$ the exponential MT-covariance of the observation vector $\Xmat$ has the following structure:
\begin{equation}
\label{MTCovExp}
\bSigma^{\left(\uexp\right)}_{\Xmatsc}\left(\tvec\right)=\Amat\bSigma^{\left(\uexp\right)}_{\Smatsc}\left(\Amat^{T}\tvec\right)\Amat^{T},
\end{equation}
where $\bSigma^{\left(\uexp\right)}_{\Smatsc}\left(\cdot\right)$ is the covariance matrix of the latent vector $\Smat$ under the transformed probability measure $Q^{\left(u_{\rm{E}}\right)}_{\Smatsc}$. Since the components of $\Smat$ are mutually independent under $\ps$, by Property \ref{P3} in Proposition \ref{Prop1}, they are mutually independent under $Q^{\left(u_{\rm{E}}\right)}_{\Smatsc}$, and therefore, $\bSigma^{\left(\uexp\right)}_{\Smatsc}\left(\cdot\right)$ must be diagonal. Thus, the following property follows directly from (\ref{MTCovExp}):
\begin{Proposition}
\label{ExpAIdent}
Let $\tvec_{1}$ and $\tvec_{2}$, $\tvec_{1}\neq\tvec_{2}$, denote two arbitrary points in $\Rsp^{p}$. Assume that
\begin{enumerate}[\upshape(1\upshape)]
\item
The matrices $\bSigma^{\left(\uexp\right)}_{\Smatsc}\left(\Amat^{T}\tvec_{1}\right)$, and $\bSigma^{\left(\uexp\right)}_{\Smatsc}\left(\Amat^{T}\tvec_{2}\right)$ have finite diagonal entries,
\item
The diagonal entries of $\bSigma^{\left(\uexp\right)}_{\Smatsc}\left(\Amat^{T}\tvec_{2}\right)$ are non-zero, and
\item
\label{A3_ExpAIdent}
The matrix $\Lambdamat^{\left(u_{\rm{E}}\right)}_{\Smatsc}\left(\Amat^{T}\tvec_{1},\Amat^{T}\tvec_{2}\right)\triangleq\bSigma^{\left(\uexp\right)}_{\Smatsc}\left(\Amat^{T}\tvec_{1}\right)\bSigma^{\left(\uexp\right)-1}_{\Smatsc}\left(\Amat^{T}\tvec_{2}\right)$ has distinct diagonal entries, i.e., no pair of diagonal entries have the same value. 
\end{enumerate}
Then, $\Amat$ can be uniquely identified, up to scaling and permutation of its columns, by solving the following non-symmetric eigenvalue decomposition problem:
\begin{equation}
\label{NSEVD}
\bSigma^{\left(\uexp\right)}_{\Xmatsc}\left(\tvec_{1}\right)\bSigma^{\left(\uexp\right)-1}_{\Xmatsc}\left(\tvec_{2}\right)\Amat=\Amat\Lambdamat^{\left(u_{\rm{E}}\right)}_{\Smatsc}\left(\Amat^{T}\tvec_{1},\Amat^{T}\tvec_{2}\right).
\end{equation}
[A proof is given in \citep{Yeredor}].
\end{Proposition}

As a result of property (\ref{ExpMTVarConst}) of the exponential MT-variance, the following Theorem shows that Assumption \ref{A3_ExpAIdent} in Proposition \ref{ExpAIdent} is satisfied almost everywhere (a.e.) if at most one of the components of $\Smat$ is Gaussian. 
\begin{Theorem}
\label{ZeroLebesgueExpTh}
If at most one of the sources is Gaussian, then for $\tvec_{1}\neq\tvec_{2}$ the matrix $\Lambdamat^{\left(u_{\rm{E}}\right)}_{\Smatsc}\left(\Amat^{T}\tvec_{1},\Amat^{T}\tvec_{2}\right)$ has distinct diagonal entries a.e. 
[A proof is given in Appendix \ref{ZeroLebesgueExpThProof}]
\end{Theorem}
\subsection{The exponential-MTICA algorithm }
According to (\ref{MTCovExp}), Proposition \ref{ExpAIdent}, and Theorem \ref{ZeroLebesgueExpTh}, the separation matrix $\Bmat=\Amat^{-1}$ is the unique matrix that jointly diagonalizes two exponential MT-covariance matrices $\bSigma^{\left(\uexp\right)}_{\Xmatsc}\left(\tvec_{1}\right)$ and $\bSigma^{\left(\uexp\right)}_{\Xmatsc}\left(\tvec_{2}\right)$ that satisfy the stated assumptions. The exponential-MTICA algorithm \citep{Yeredor} is obtained by replacing the MT-functions $u_{m}\left(\cdot\right)$, $m=1,\ldots,M$, in Algorithm \ref{EMTICAAlg} with a sequence of exponential MT-functions $\uexp\left(\cdot;\tvec_{m}\right)$, $m=1,\ldots,M$. A procedure for choosing the test-points $\tvec_{m}\in\Rsp^{p}$, $m=1,\ldots,M$, is given in Appendix \ref{EMTICA_PARAM}. Clearly, only two test-points are needed for obtaining a viable estimate of $\Bmat$. However, in order to increase statistical stability and reduce the effect of ill-conditioned empirical MT-covariance matrices it is better to use a sequence of more than two test-points. 

\section{Gaussian-MTICA}
\label{GMTICA}
In this section we parameterize the MT-function $u\left(\cdot;\tvec,\tau\right)$, with translation parameter $\tvec\in\Rsp^{p}$ and width parameter $\tau\in\Rsp^{*}_{+}$
using a Gaussian family of functions. Under the unitary mixing model (\ref{ZDef}) we show that if at most one of the sources is Gaussian, the mixing matrix $\Umat$ can be uniquely identified via eigenvalue decomposition of a single Gaussian MT-covariance matrix. Based on this result the Gaussian-MTICA algorithm is obtained that applies OAJD to a sequence of empirical Gaussian MT-covariance matrices.
\subsection{The Gaussian MT-covariance}
\label{GaussMTCov}
We define the Gaussian MT-function $u_{\rm{G}}\left(\cdot;\cdot,\cdot\right)$ as 
\begin{eqnarray}
\label{GaussKernel}  
\uGausss\left(\xvec;\tvec,\tau\right)\triangleq\exp\left(-\frac{\left\|\xvec-\tvec\right\|^{2}_{2}}{2\tau^{2}}\right),
\end{eqnarray}
where $\tvec\in\Rsp^{p}$, and $\tau\in\Rsp^{*}_{+}$. Since $\uGausss\left(\cdot;\cdot,\cdot\right)$ is strictly positive and bounded, one can easily verify that the condition (\ref{Assumption2}) is always satisfied. Relations (\ref{VarPhiDef}) and (\ref{MTCovZ}) imply that the MT-function (\ref{GaussKernel}) produces a weighted covariance matrix, $\bSigma^{\left(\uGausss\right)}_{\Xmatsc}\left(\tvec,\tau\right)$, for which the observations are weighted in inverse proportion to the distance $\left\|\xvec-\tvec\right\|^{2}_{2}$. This results in a kind of local covariance analysis of $\Xmat$ in the vicinity of the test-point $\tvec$. Notice that the Gaussian MT-function (\ref{GaussKernel}) satisfies the conditions in Proposition \ref{RobustnessConditions}, and therefore, the influence function of the empirical Gaussian MT-covariance is bounded. Hence, unlike the empirical exponential MT-covariance, whose influence function is unbounded, the empirical Gaussian MT-covariance is robust to outlying observations. 

Similarly to the exponential measure transformation, the following lemma, directly following from (\ref{VarPhiDef})-(\ref{RadNik2}) and the definition of the Gaussian MT-function (\ref{GaussKernel}), states that the Gaussian family of probability measures is closed under the measure transformation (\ref{MeasureTransform}) generated by the Gaussian MT-function. 
\begin{Lemma}
\label{GaussPresProp}
A random vector $\Xmat$ is Gaussian under the probability measure $\px$ if and only if it remains Gaussian under the transformed probability measure $Q^{\left(\uGausss\right)}_{\Xmatsc}$ generated by the Gaussian MT-function $\uGausss\left(\cdot;\cdot,\cdot\right)$.
\end{Lemma}
This property is used in proving the following theorem that states a necessary and sufficient condition for Gaussianity of a random variable $X$ based on its Gaussian MT-variance.
\begin{Theorem}
\label{ConstVarGaussMT}
A random variable $X$ with corresponding probability measure $P_{X}$ is Gaussian if and only if the the Gaussian MT-variance satisfies
\begin{equation}
\label{GaussMTVarConst}
\sigma^{\left(\uGausss\right)}_{X}\left(t,\tau\right)=c\hspace{0.2cm}\forall{t}\in\left(t_{0}-\epsilon,t_{0}+\epsilon\right),
\end{equation}
where $c$ and $\epsilon$ are some positive constants and $t_{0}$ is some arbitrary point in $\Rsp$ [A proof is given in Appendix \ref{ConstVarGaussMTProof}].
\end{Theorem}
Hence, similarly to the exponential MT-variance, if a random variable $X$ is non-Gaussian then for any choice of the width parameter $\tau\in\Rsp^{*}_{+}$ the Gaussian MT-variance $\sigma^{\left(\uGausss\right)}_{X}\left(t,\tau\right)$ cannot be constant w.r.t. $t$ over any open interval. This property is used in the following subsection for proving identifiability of the mixing matrix $\Umat$. 
\subsection{Identifiability of the unitary mixing matrix $\Umat$ under a single Gaussian MT-covariance}
According to (\ref{ZDef}), (\ref{VarPhiDef}), (\ref{MTCovZ}) and (\ref{GaussKernel}) the MT-covariance of the whitened observation vector $\Zmat$ under $Q^{(\uGausss)}_{\Zmatsc}$ 
has the following structure:
\begin{equation}
\label{MTCovGauss}  
\bSigma^{\left(\uGausss\right)}_{\Zmatsc}\left(\tvec,\tau\right)=\Umat\bSigma^{\left(\uGausss\right)}_{\Smatsc}\left(\Umat^{T}\tvec,\tau\right)\Umat^{T},
\end{equation}
where $\bSigma^{\left(\uGausss\right)}_{\Smatsc}\left(\cdot,\cdot\right)$ is the covariance matrix of $\Smat$ under the transformed probability measure $Q^{(\uGausss)}_{\Smatsc}$. Since the components of $\Smat$ are mutually independent under $\ps$, then by Property \ref{P3} in Proposition \ref{Prop1} they are mutually independent under $Q^{(\uGausss)}_{\Smatsc}$, and thus, $\bSigma^{\left(\uGausss\right)}_{\Smatsc}\left(\cdot,\cdot\right)$ must be diagonal. Therefore, assuming that $\bSigma^{\left(\uGausss\right)}_{\Smatsc}\left(\Umat^{T}\tvec,\tau\right)$ has distinct finite diagonal entries, the unitary matrix $\Umat$ can be uniquely identified (up to permutation and sign of its columns) via eigenvalue decomposition of the Gaussian MT-covariance $\bSigma^{\left(\uGausss\right)}_{\Zmatsc}\left(\tvec,\tau\right)$. 

Based on property (\ref{GaussMTVarConst}) of the Gaussian MT-variance, the following theorem states that if at most one of the components of $\Smat$ is Gaussian, then $\bSigma^{\left(\uGausss\right)}_{\Smatsc}\left(\Umat^{T}\tvec,\tau\right)$ has distinct diagonal entries for almost every $\tvec\in\Rsp^{p}$.
\begin{Theorem}
\label{ZeroLebesgueGauss}
If at most one of the sources is Gaussian, then the matrix $\bSigma^{\left(\uGausss\right)}_{\Smatsc}\left(\Umat^{T}\tvec,\tau\right)$ has distinct diagonal entries a.e.
[A proof is given in Appendix \ref{ZeroLebesgueGaussProof}]
\end{Theorem}
\subsection{The Gaussian-MTICA algorithm}
According to (\ref{MTCovGauss}) and Theorem \ref{ZeroLebesgueGauss}, if at most one of the sources is Gaussian, then for almost every $\tvec\in\Rsp^{p}$ the matrix $\Vmat=\Umat^{T}$ is the unique diagonalizing matrix of the Gaussian MT-covariance $\bSigma^{\left(\uGausss\right)}_{\Zmatsc}\left(\tvec,\tau\right)$. Thus, the Gaussian-MTICA algorithm is implemented by replacing the MT-functions $u_{m}\left(\cdot\right)$, $m=1,\ldots,M$ in Algorithm \ref{GMTICAAlg} with Gaussian MT-functions $\uGausss\left(\cdot;\tvec_{m},\tau\right)$, $m=1,\ldots,M$, where the width parameter $\tau\in\Rsp^{*}_{+}$ is fixed. A procedure for choosing the test-points $\tvec_{m}\in\Rsp^{p}$, $m=1,\ldots,M$ is given in Appendix \ref{GMTICA_PARAM}. Clearly, only one test-point is needed for estimating $\Vmat$. However, estimation of $\Vmat$ based on diagonalization of a single empirical Gaussian MT-covariance has the following drawbacks:
\begin{inparaenum}[\upshape(1\upshape)] 
\item
For some choice of the translation parameter $\tvec$ the eigen-spectrum of the corresponding Gaussian MT-covariance may be degenerate, i.e., the eigenvalues may not be well separated. 
\item
A single Gaussian MT-covariance may only capture part of the statistical information about $\Zmat$ necessary to separate the sources effectively.
\end{inparaenum}
In order to alleviate these drawbacks it may be better to use more than a single test-point. 
\section{Comparisons between exponential and Gaussian MTICA}
\label{EMTICAvsGMTICA}
Unlike Gaussian-MTICA that requires whitening, which under the model  (\ref{ICAModel}) leads to unitary mixing, exponential-MTICA does not require whitening. Therefore, as illustrated in Subsection \ref{ModelMismatch}, exponential-MTICA is more robust to cases where the whitened observations are poorly modeled by unitary mixing. Moreover, in Gaussian-MTICA, one has to set a width parameter $\tau$ not required for exponential-MTICA. 

On the other hand, unlike the exponential MT-function, the Gaussian MT-function is bounded and isotropically de-emphasizes samples distant from its location parameter. This property leads to the following advantages of Gaussian-MTICA over exponential-MTICA: 
\begin{inparaenum}[\upshape(1\upshape)]
\item
As illustrated in Subsections \ref{SourceDist} and \ref{Outliers}, Gaussian-MTICA is more robust to heavy-tailed distributions and outliers than exponential-MTICA. 
\item
Unlike the exponential MT-covariance, which does not exist for distributions with infinite moment generating function, the Gaussian MT-covariance takes finite values regardless of the underlying probability distribution. 
\end{inparaenum}
Additionally, the Gaussian MT-function has the physical property that it localizes linear dependence over the observation space. Hence, Gaussian-MTICA operates by jointly minimizing the local linear dependencies in the vicinities of the selected set of test-points. 
\section{Computational complexity}
\label{ACL}
The exponential-MTICA has two major steps: 
\begin{inparaenum}[\upshape(1\upshape)]
\item
estimation of $M$ exponential MT-covariance matrices with computational complexity of $O\left({M}\cdot{N}\cdot{p^{2}}\right)$ flops; and
\item
NOAJD. 
\end{inparaenum}
The computational complexity of unweighted NOAJD algorithms, such as Pham's \citep{Pham}, FFDIAG \citep{FFDIAG}, QDIAG \citep{QDIAG}, and U-WEDGE \citep{FastJDWM} is $O\left(L\cdot{M}\cdot{p}^{3}\right)$ flops, where $L$ is the number of iterations. Hence, exponential-MTICA with unweighted NOAJD has computational complexity of $O\left({M}\cdot{N}\cdot{p^{2}}+L\cdot{M}\cdot{p}^{3}\right)$ flops. When weighted NOAJD is applied using the WEDGE algorithm \citep{FastJDWM} with the weighting policy proposed in \citep{WITCHESS}, one has to calculate the weights, with computational complexity of $O\left({M}^{2}\cdot{N}\cdot{p}+{M}^{2}\cdot{p}^{2}\right)$ flops, and to apply weighted NOAJD requiring $O\left(L\cdot{M}\cdot{p}^{3}\right)$ flops. Therefore, exponential-MTICA with weighted NOAJD requires $O\left({M}^{2}\cdot{N}\cdot{p}+({M}^{2}+{M}\cdot{N})\cdot{p^{2}}+L\cdot{M}\cdot{p}^{3}\right)$ flops.

The Gaussian-MTICA algorithm has three steps: 
\begin{inparaenum}[\upshape(1\upshape)]
\item
a whitening stage with computational complexity of $O\left({N}\cdot{p}^{2}\right)$ flops,
\item
estimation of $M$ Gaussian MT-covariance matrices with computational complexity of $O\left({M}\cdot{N}\cdot{p^{2}}\right)$ flops, and
\item
OAJD with computational complexity of $O\left(L\cdot{M}\cdot{p}^{3}\right)$ flops \citep{FG, EJM}.
\end{inparaenum}
Thus, Gaussian-MTICA requires $O\left({M}\cdot{N}\cdot{p^{2}}+L\cdot{M}\cdot{p}^{3}\right)$ flops. 

Table \ref{ACLTAB} compares the computational complexity of exponential-MTICA and Gaussian-MTICA to the computational complexity of other ICA techniques, such as JADE, EIMAX, FICA, EFICA, KGV, and RADICAL. Notice that similarly to JADE, FICA, EFICA, and EIMAX the computational complexities of exponential-MTICA and Gaussian-MTICA are linear in the sample size $N$, which make them favorable for large data sets. Moreover, one sees that unlike data-based techniques such as EIMAX, FICA, EFICA, KGV, and RADICAL, the iterative part of exponential-MTICA and Gaussian-MTICA has computational complexity that is not affected by the sample size. 
\begin{table}[htdp]
\caption{Computational complexity of EMTICA, GMTICA, JADE, EIMAX, FICA, EFICA, KGV, and RADICAL.
The samples size, dimension, number of iterations, and number of matrices to be approximately diagonalized are denoted by $N$, $p$, $L$, and $M$, respectively. The rank of an $N\times{N}$ Gram matrix after incomplete Cholesky decomposition in the KGV is denoted by $D(N)$. The number of Jacobi angles, and data augmentations in RADICAL are denoted by $K$ and $R$, respectively. Here EMTICA and GMTICA refer to exponential-MTICA and Gaussian-MTICA, respectively. NOAJD stands for non-orthogonal joint diagonalization.}
\begin{center}
\begin{tabular}{| l | l |}
\hline
\hspace{0cm}\textbf{Algorithm}   & \hspace{0cm}\textbf{Computational complexity} \\
\hline
\hspace{0cm}EMTICA  (unweighted NOAD) & \hspace{0cm}{$O\left({M}\cdot{N}\cdot{p^{2}}+L\cdot{M}\cdot{p}^{3}\right)$}. \\
\hline
\hspace{0cm}EMTICA  (weighted NOAD) & \hspace{0cm}{$O\left({M}^{2}\cdot{N}\cdot{p}+({M}^{2}+{M}\cdot{N})\cdot{p^{2}}+L\cdot{M}\cdot{p}^{3}\right)$}. \\
\hline
\hspace{0cm}GMTICA & \hspace{0cm}{$O\left({M}\cdot{N}\cdot{p^{2}}+L\cdot{M}\cdot{p}^{3}\right)$}. \\
\hline
\hspace{0cm}JADE & \hspace{0cm}{$O\left({M}\cdot{N}\cdot{p^{2}}+L\cdot{M}\cdot{p}^{3}\right)$}. \\
\hline
\hspace{0cm}EIMAX & \hspace{0cm}{$O\left(L\cdot{N}\cdot{p}^{3}\right)$}. \\
\hline
\hspace{0cm}FICA & \hspace{0cm}{$O\left(L\cdot{N}\cdot{p^{2}}\right)$}. \\
\hline
\hspace{0cm}EFICA & \hspace{0cm}{$O\left(L\cdot{N}\cdot{p^{2}}\right)$}. \\
\hline
\hspace{0cm}KGV  & \hspace{0cm}{$O\left(L\cdot\left({N}\cdot{D^{2}(N)}\cdot{p}^{2}+{D}^{3}(N)\cdot{p}^{3}\right)\right)$.}\\
\hline
\hspace{0cm}RADICAL & \hspace{0cm}{$O\left(L\cdot\left(K\cdot{N}\cdot{R}\cdot\log\left(N\cdot{R}\right)\cdot{p}^{2}\right)\right)$. } \\
\hline
\end{tabular}
\end{center}
\label{ACLTAB}
\end{table}
\section{Numerical examples}
\label{Examp}
In this Section, the performances of exponential-MTICA and Gaussian-MTICA are compared to the JADE, EIMAX, FICA, EFICA, KGV, and RADICAL algorithms using their publicly available MATLAB code. The JADE, FICA, EFICA, EIMAX, and RADICAL algorithms were used with their default settings. In KGV the Gaussian kernel width parameter was set to $\sigma=1$. All compared algorithms were initialized by the identity separation matrix.

The test-points $\tvec_{1},\ldots,\tvec_{M}$ in the exponential and Gaussian MTICA algorithms were selected according to the procedures in Appendices \ref{EMTICA_PARAM} and \ref{GMTICA_PARAM}, respectively. In all simulation examples $M=30$ test-points were used. The width parameter of the Gaussian MT-function in the Gaussian-MTICA algorithm was set to $\tau=1$. 

The exponential-MTICA was implemented with the WEDGE algorithm \citep{FastJDWM} to perform weighted non-orthogonal joint diagonalization as proposed in \citep{WITCHESS}.
The Gaussian-MTICA was implemented with the FG algorithm \citep{FG} to perform orthogonal joint diagonalization. In all considered approximate joint diagonalization algorithms, the initial diagonalizing matrix, the maximum number of iterations and the convergence threshold were set to the identity matrix, 500 and 1e-10, respectively. In all figure legends below, the exponential and Gaussian MTICA algorithms are abbreviated by EMTICA and GMTICA, respectively. 

We used the Amari error \citep{Amari} to measure the deviation of the true separation matrix $\Bmat$ from its estimate $\hat{\Bmat}$. The Amari error between two matrices $\Gmat\in\Rsp^{p\times{p}}$ and $\Hmat\in\Rsp^{p\times{p}}$ is defined as:
\begin{equation}
d_{\rm{A}}\left(\Gmat,\Hmat\right) = \frac{1}{2p(p-1)}\sum\limits_{i=1}^{p}\left(\frac{\sum_{j=1}^{p}\left|\Psi_{i,j}\right|}{\max_{j}\left|\Psi_{i,j}\right|}-1\right)
+\frac{1}{2p(p-1)}\sum\limits_{j=1}^{p}\left(\frac{\sum_{i=1}^{p}\left|\Psi_{i,j}\right|}{\max_{i}\left|\Psi_{i,j}\right|}-1\right),
\end{equation}
where $\Psi_{i,j}=\left[\Gmat\Hmat^{-1}\right]_{i,j}$. The Amari error is invariant to permutation and scaling of the columns of $\Gmat$ and $\Hmat$, and takes values between 0 and 1. Another property is that $d_{\rm{A}}\left(\Gmat,\Hmat\right)=0$ if and only if $\Gmat$ and $\Hmat$ are equal up to scaling and permutation of their columns. In addition to the Amari error, some of the trials compared the algorithm run times. 

The simulations were carried out using data obtained from the univariate source distributions in Table \ref{DistributionsTable}.
\begin{table}[htdp]
\caption{Probability distributions used in the simulation examples.}
\begin{center}
\begin{tabular}{| l | l |}
\hline
\hspace{0cm}\textbf{Distribution}   & \hspace{0cm}\textbf{Parameters} \\
\hline
\hspace{0cm}Uniform & \hspace{0cm}{Support $\left[0,1\right]$}. \\
\hline
\hspace{0cm}Arcsine & \hspace{0cm}{Support $\left[0,1\right]$}. \\
\hline
\hspace{0cm}Laplace & \hspace{0cm}{Location parameter $\mu=0$ and scale parameter $\sigma=1$}. \\
\hline
\hspace{0cm}Student t-distribution & \hspace{0cm}{Degrees of freedom $\kappa=3$}. \\
\hline
\hspace{0cm}Beta & \hspace{0cm}{Shape parameters $\alpha=2$ and $\beta=2$}. \\
\hline
\hspace{0cm}Exponential & \hspace{0cm}{Rate parameter $\lambda=1$}. \\
\hline
\hspace{0cm}Rayleigh & \hspace{0cm}{Scale parameter $\sigma=1$}. \\
\hline
\hspace{0cm}Gamma & \hspace{0cm}{Shape parameter $\alpha=1$ and scale parameter $\sigma=1$}. \\
\hline
\hspace{0cm}Central chi-squared& \hspace{0cm}{Degrees of freedom $\kappa=4$}. \\
\hline
\hspace{0cm}Rice & \hspace{0cm}{Shape parameter $\alpha=1/2$}. \\
\hline
\end{tabular}
\end{center}
\label{DistributionsTable}
\end{table}
The sources were translated and scaled to have zero mean and unit variance. In order to avoid ill-conditioned mixing, the generated sources were mixed using randomly generated matrices having condition number between one and two. In all experiments we studies 5-dimensional ICA problems with $N=1000$ samples.
\subsection{Sensitivity to source distribution}
\label{SourceDist}
\begin{figure}[htbp!]
  \begin{center}
    {{\subfigure{\label{GraphModel_3_MTCCA_EXP_1}\includegraphics[scale=0.286]{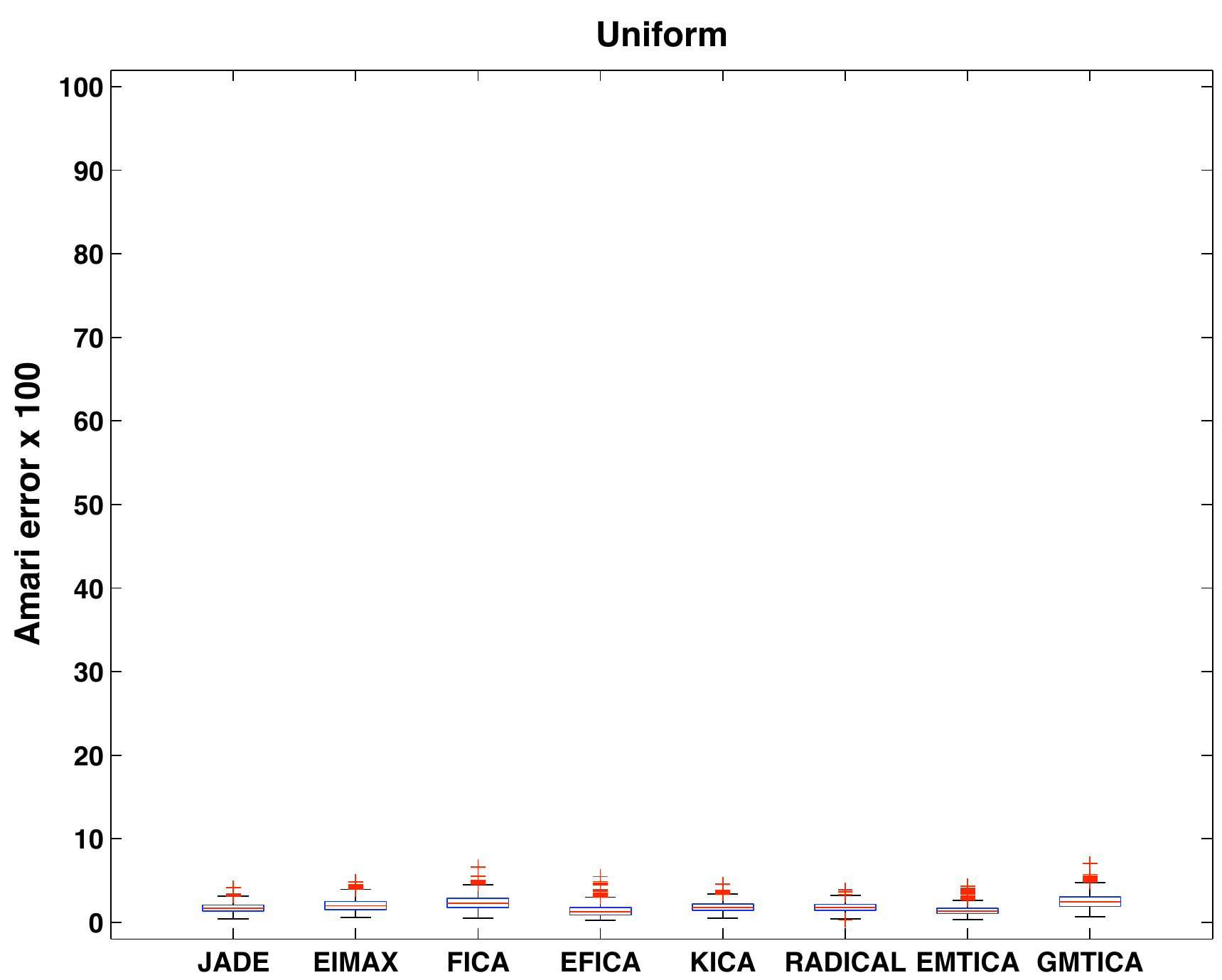}}}}
    {{\subfigure{\label{GraphModel_3_MTCCA_EXP_2}\includegraphics[scale=0.286]{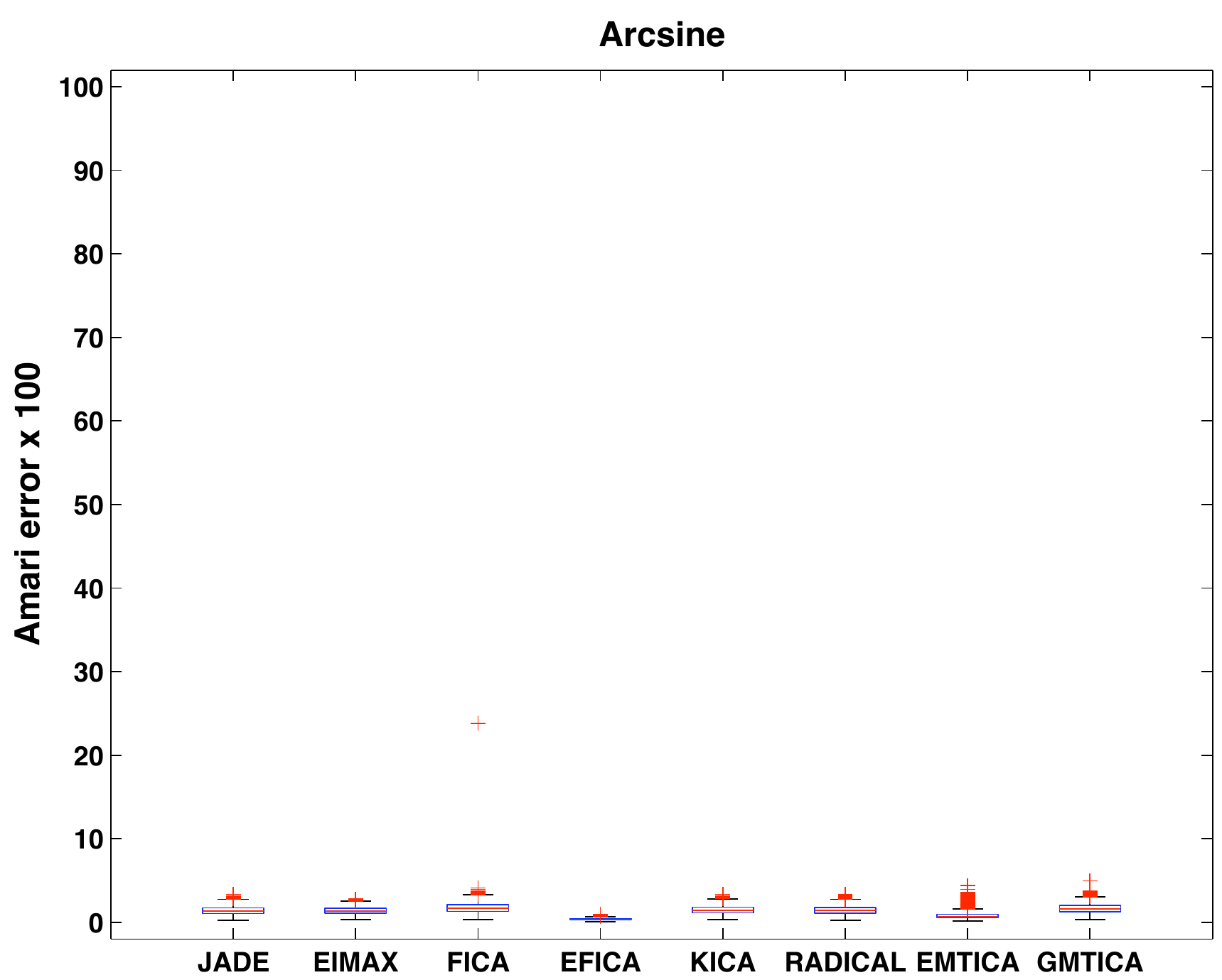}}}}
    {{\subfigure{\label{GraphModel_3_MTCCA_EXP_2}\includegraphics[scale=0.286]{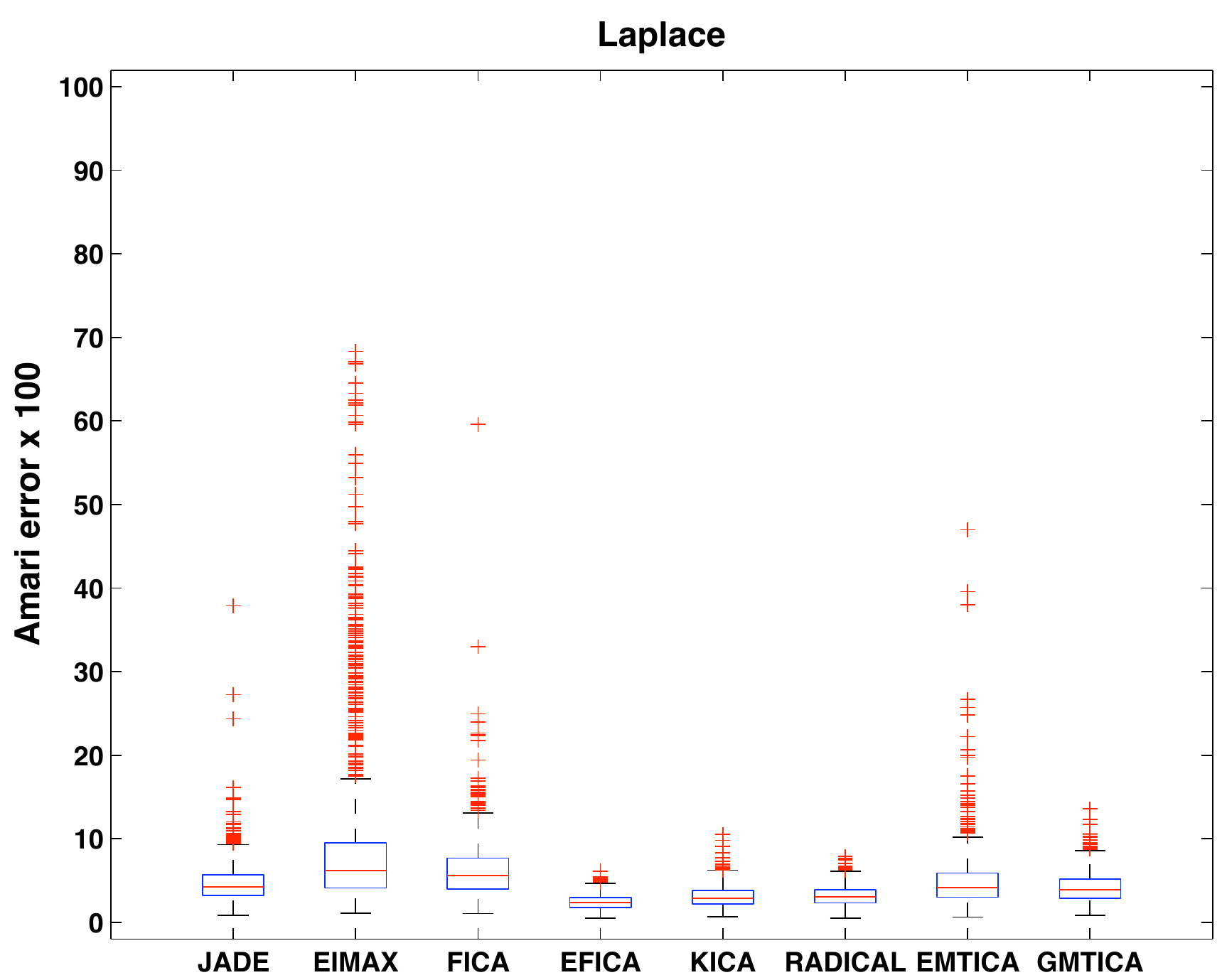}}}}
    {{\subfigure{\label{GraphModel_3_MTCCA_EXP_3}\includegraphics[scale=0.286]{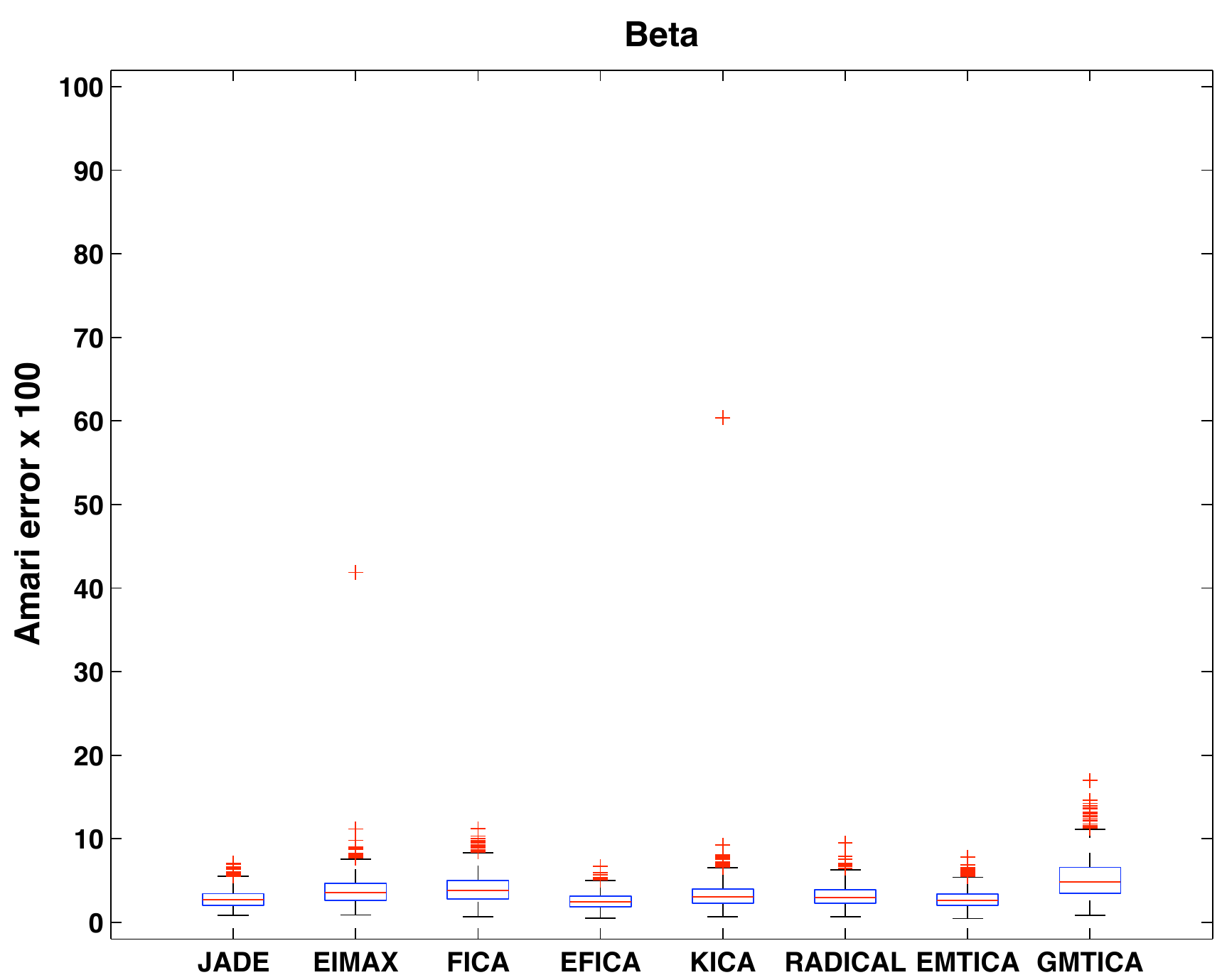}}}}
    {{\subfigure{\label{GraphModel_3_MTCCA_EXP_5}\includegraphics[scale=0.286]{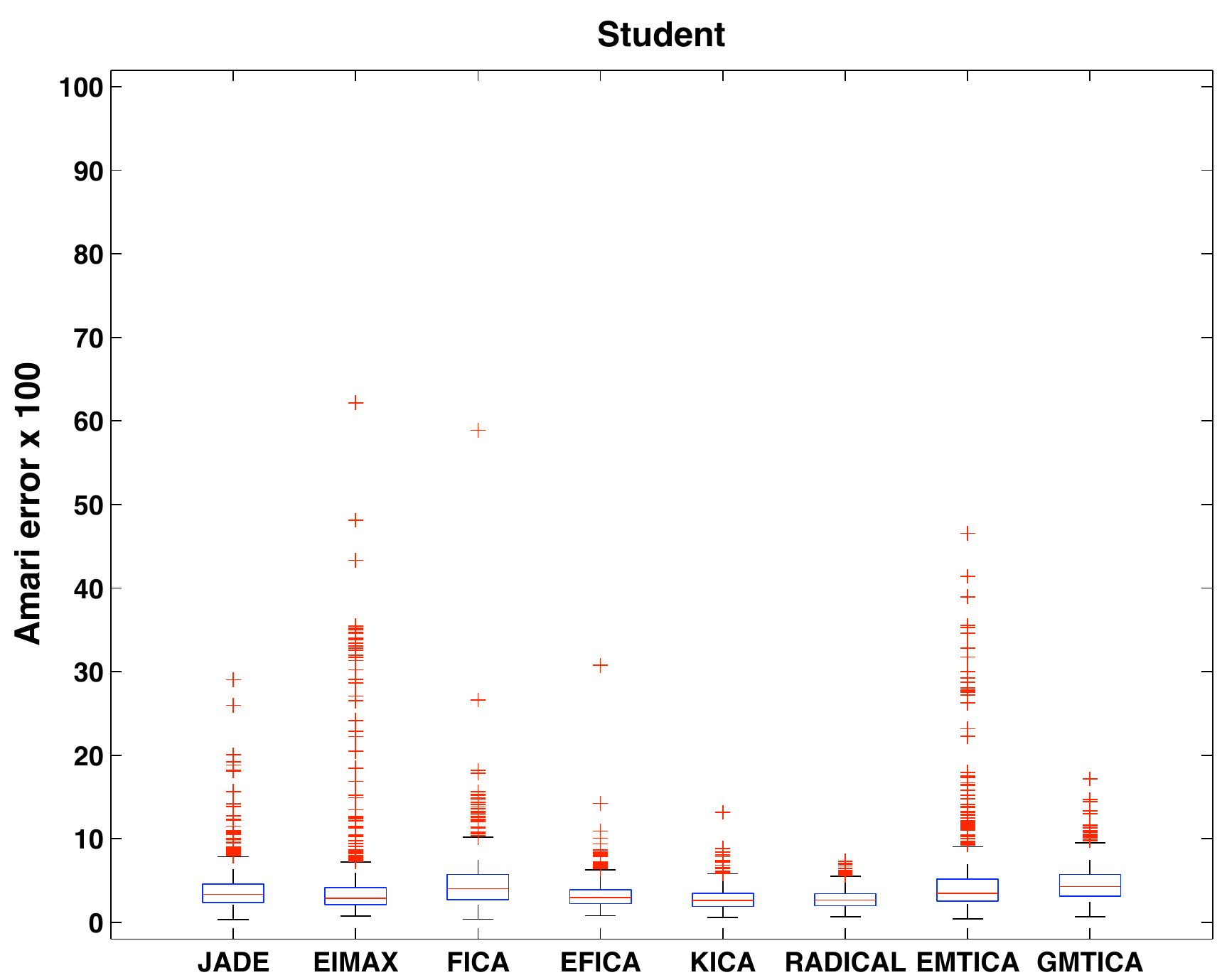}}}}
    {{\subfigure{\label{GraphModel_3_MTCCA_EXP_6}\includegraphics[scale=0.286]{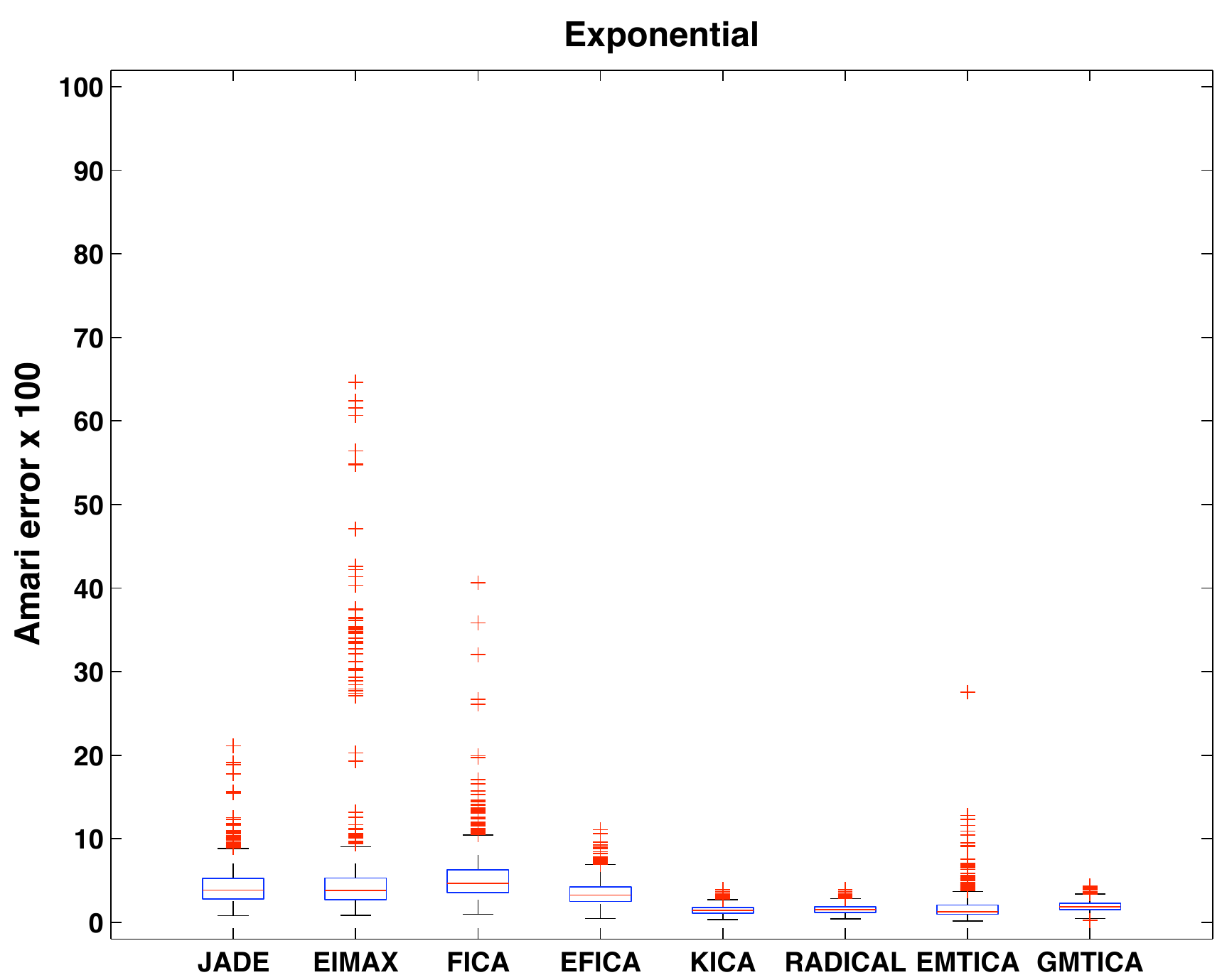}}}}
    {{\subfigure{\label{GraphModel_3_MTCCA_EXP_7}\includegraphics[scale=0.286]{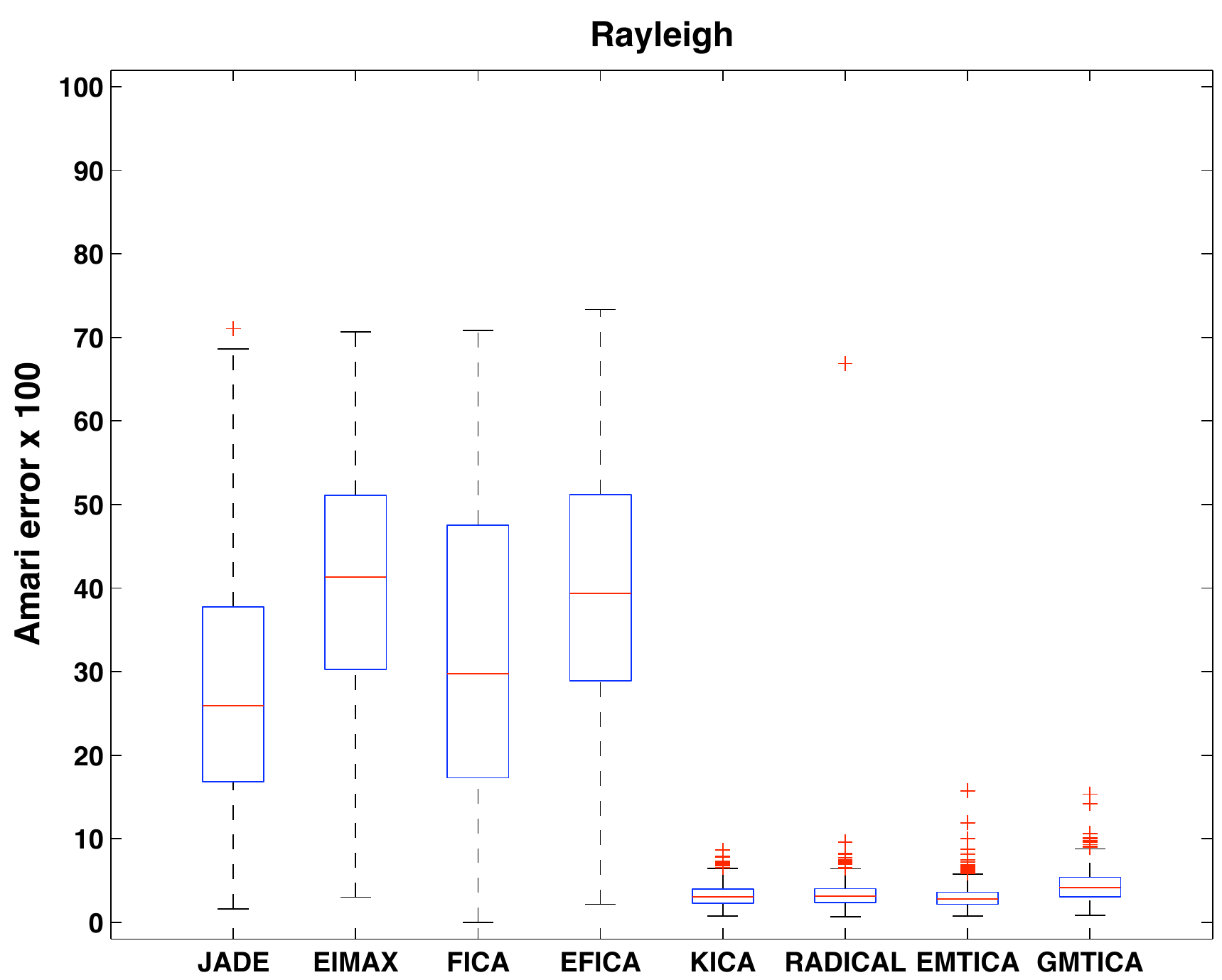}}}}
    {{\subfigure{\label{GraphModel_3_MTCCA_EXP_9}\includegraphics[scale=0.286]{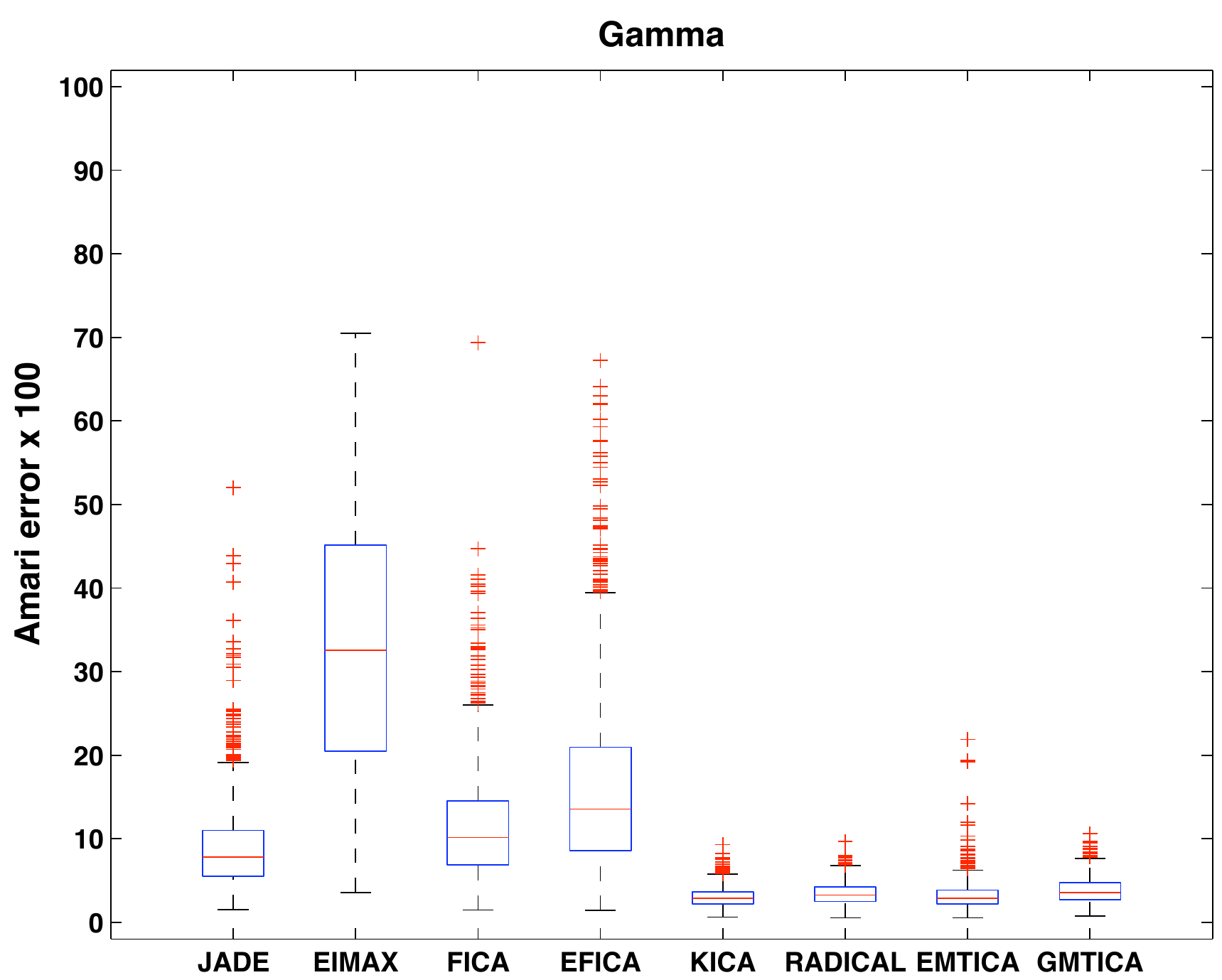}}}}
    {{\subfigure{\label{GraphModel_3_MTCCA_EXP_9}\includegraphics[scale=0.286]{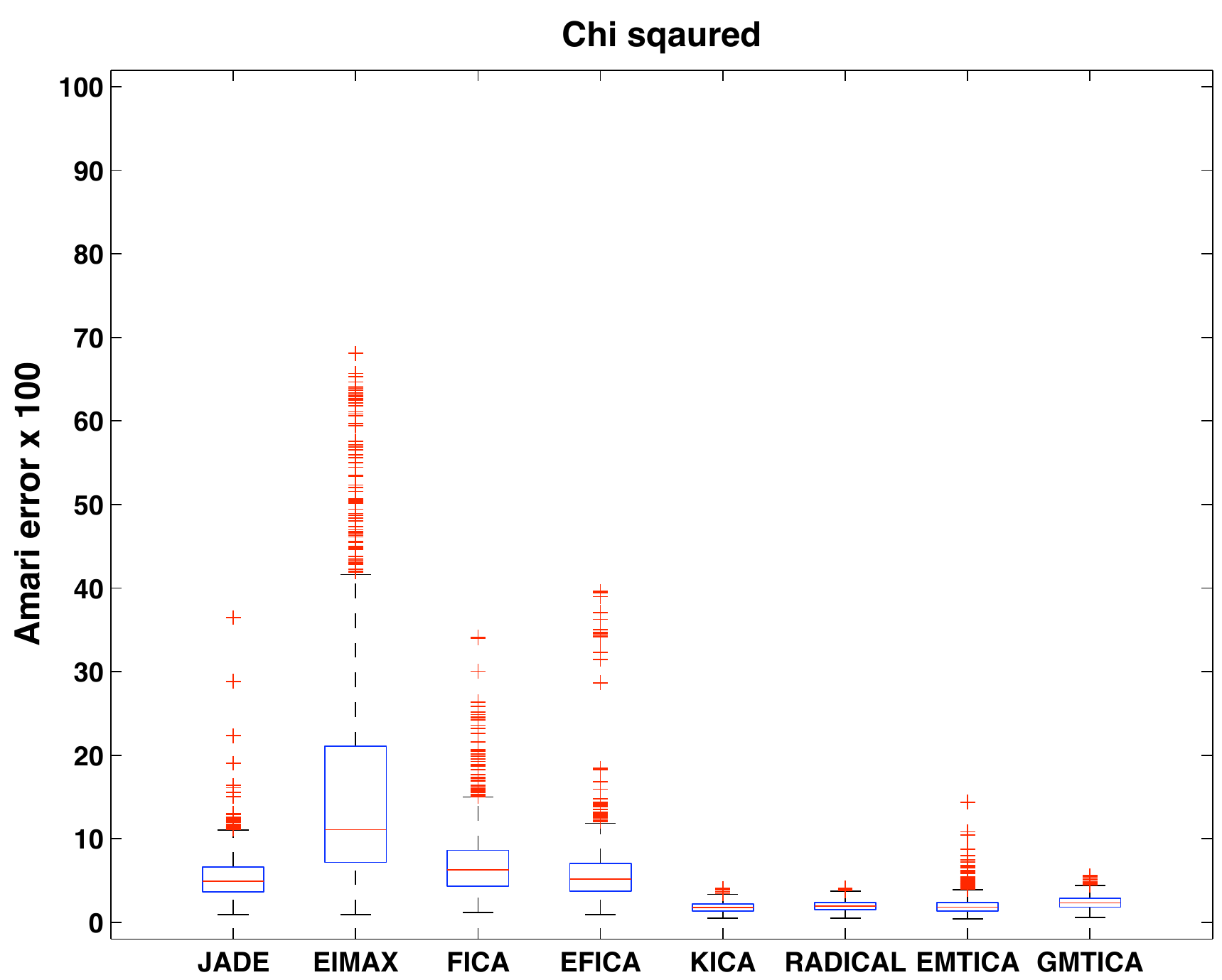}}}}
    {{\subfigure{\label{GraphModel_3_MTCCA_EXP_9}\includegraphics[scale=0.286]{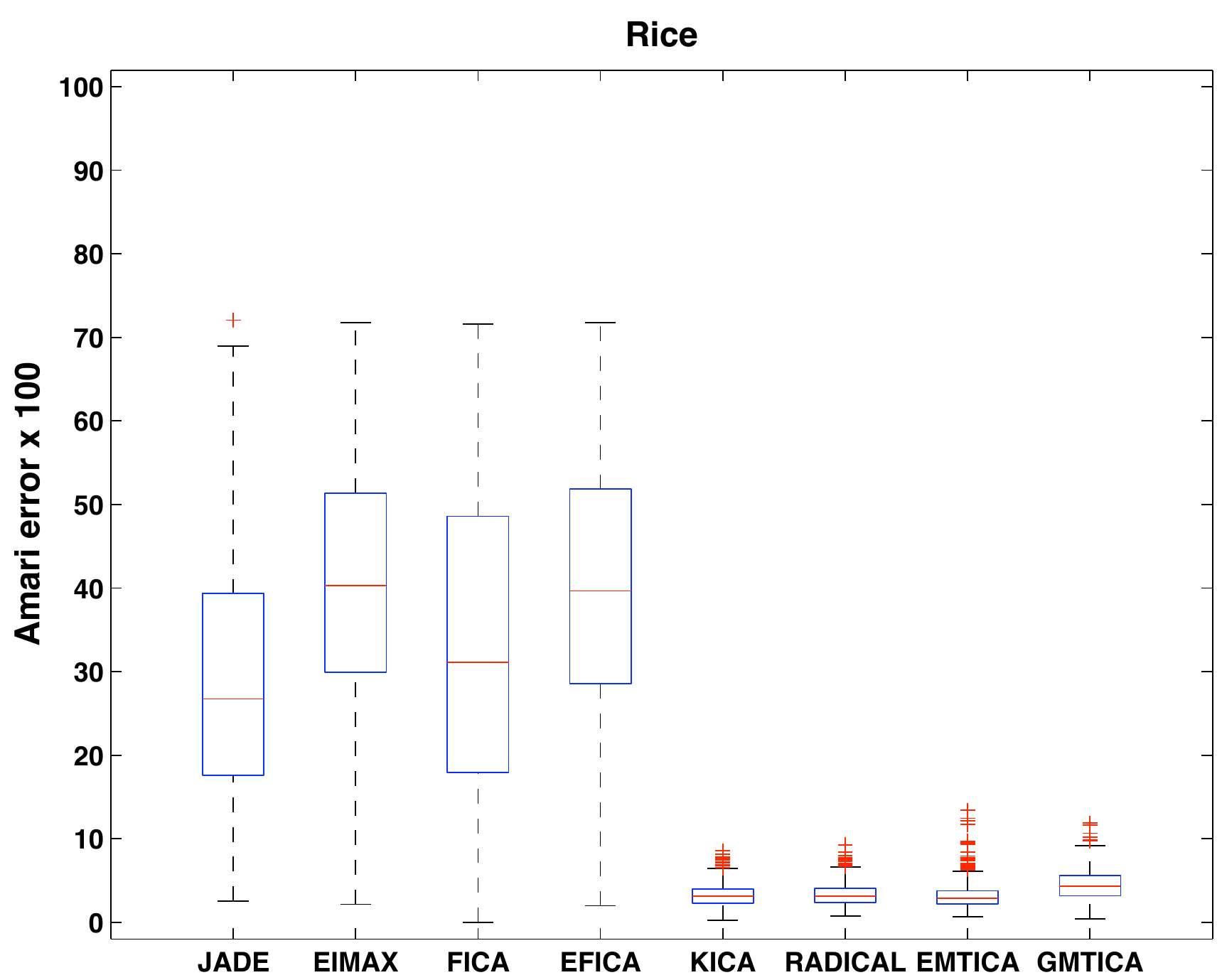}}}}
  \end{center}
  \caption{Sensitivity to source distribution. Box plots of Amari errors obtained by the compared algorithms for five-component ICA with identical source distributions. Notice that exponential-MTICA and Gaussian-MTICA are robust to source distribution with performance similar to the KGV and RADICAL algorithms. Although the exponential-MTICA, Gaussian-MTICA, KGV and RADICAL algorithms perform similarly well, the exponential-MTICA and Gaussian-MTICA have reduced computational complexity as indicated by Table \ref{ACLTAB}. Also notice that Gaussian-MTICA is less sensitive to heavy-tailed distributions than exponential-MTICA.}
\label{Fig1}
\end{figure}

In this experiment we studied two types of ICA applications. In the first application, the source distributions are identical. For each of the 10 source distributions in Table \ref{DistributionsTable}, we conducted 1000 Monte-Carlo simulations. For each distribution type, box plots of the Amari errors obtained by each algorithm are depicted in Fig. \ref{Fig1}. One sees that exponential-MTICA and Gaussian-MTICA are more robust to source distribution than JADE, EIMAX, FICA and EFICA, with performance similar to the KGV and RADICAL algorithms. One can also observe that exponential-MTICA is more sensitive to heavy-tailed distributions, such as Laplace, exponential, and student-t than Gaussian-MTICA. 

\begin{figure}[htbp!]
  \begin{center}
    {{{\label{GraphModel_3_MTCCA_EXP_1}\includegraphics[scale=0.7]{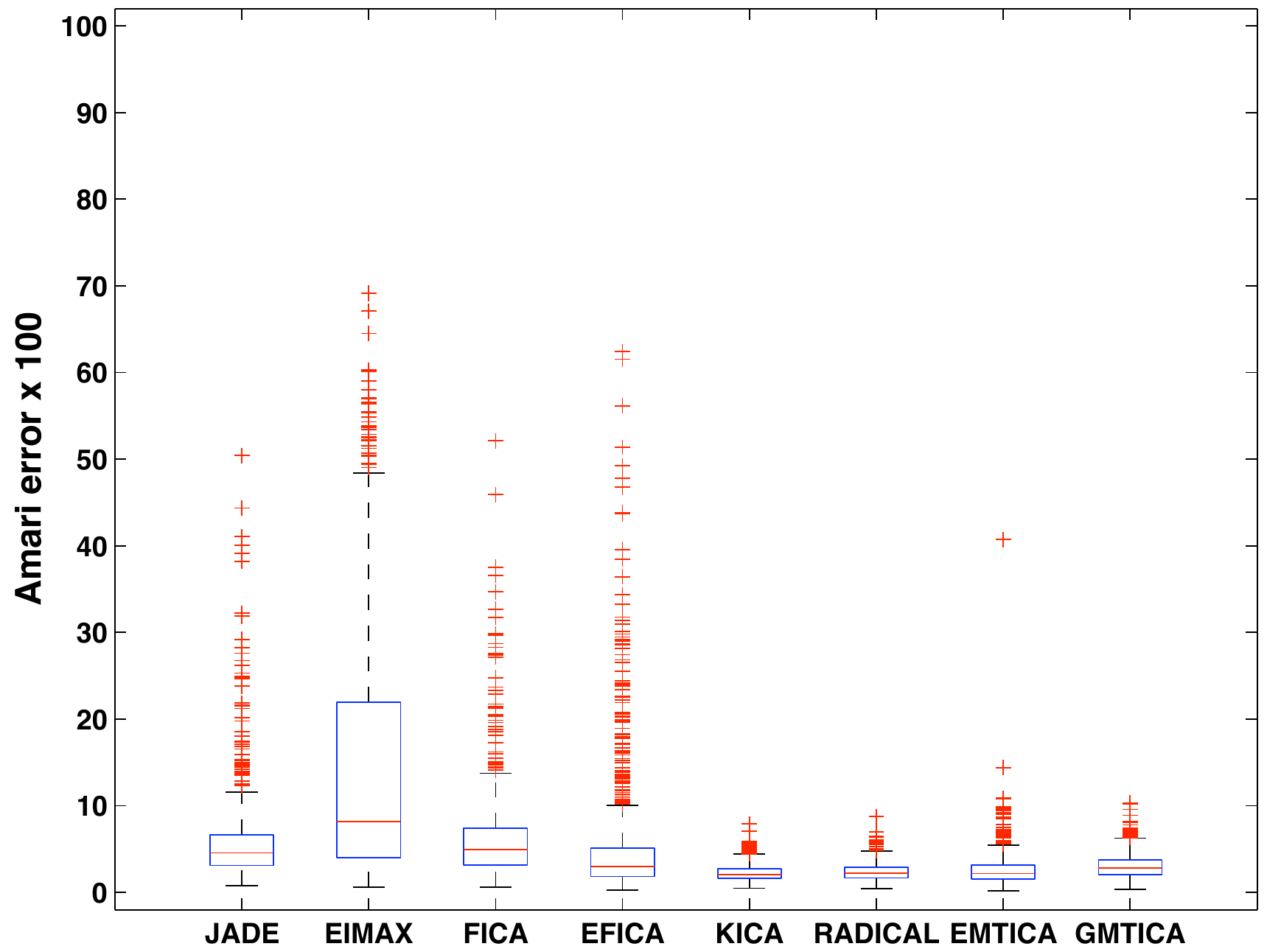}}}}
  \end{center}
  \caption{Sensitivity to source distribution. Box plots of Amari errors obtained by the compared algorithms for five-component ICA with randomly chosen distributions.
  In similar to the KGV and RADICAL, the exponential-MTICA and Gaussian-MTICA perform better than JADE, FICA, EFICA, and EIMAX algorithms. Although the exponential-MTICA, Gaussian-MTICA, KGV, and RADICAL algorithms perform similarly well, the exponential-MTICA and Gaussian-MTICA have reduced computational complexity as indicated by Table \ref{ACLTAB} and the run time analysis in Table \ref{RunTimeTab}.}
\label{Fig2}
\end{figure}

In the second application, the sources were randomly chosen among the 10 possibilities. A total of 1000 Monte-Carlo simulations were performed. The box plots of the Amari errors obtained by each algorithm are depicted in Fig. \ref{Fig2}. Notice that, similarly to the KGV and RADICAL, the exponential-MTICA and Gaussian-MTICA performs better than JADE, FICA, EFICA and EIMAX algorithms. The average run time of each algorithm is given in Table \ref{RunTimeTab}. The run times of exponential-MTICA and Gaussian-MTICA are significantly lower than those obtained by KGV and RADICAL. This is due to lower computational complexity, as indicated by Table \ref{ACLTAB}, and more rapid convergence. 
\begin{table}[htdp]
\caption{Sensitivity to source distribution. Average run times in seconds for five-component ICA problems with randomly selected source distributions. One sees that the run times of exponential-MTICA and Gaussian-MTICA are significantly lower than those obtained by KGV and RADICAL algorithms.}
\begin{center}
\begin{tabular}{| l | l |}
\hline
\hspace{0cm}\textbf{Algorithm}   & \hspace{0cm}\textbf{Run time [sec]} \\
\hline
\hspace{0cm}EMTICA & \hspace{0cm}{$3$} \\
\hline
\hspace{0cm}GMTICA & \hspace{0cm}{$0.1$} \\
\hline
\hspace{0cm}JADE & \hspace{0cm}{$0.01$} \\
\hline
\hspace{0cm}EIMAX & \hspace{0cm}{$1$} \\
\hline
\hspace{0cm}FICA & \hspace{0cm}{$0.04$} \\
\hline
\hspace{0cm}EFICA & \hspace{0cm}{$0.08$} \\
\hline
\hspace{0cm}KGV  & \hspace{0cm}{$9$} \\
\hline
\hspace{0cm}RADICAL & \hspace{0cm}{${58}$} \\
\hline
\end{tabular}
\end{center}
\label{RunTimeTab}
\end{table}
\subsection{Robustness to outliers}
\label{Outliers} 
\begin{figure}[htbp!]
\centerline{\psfig{figure=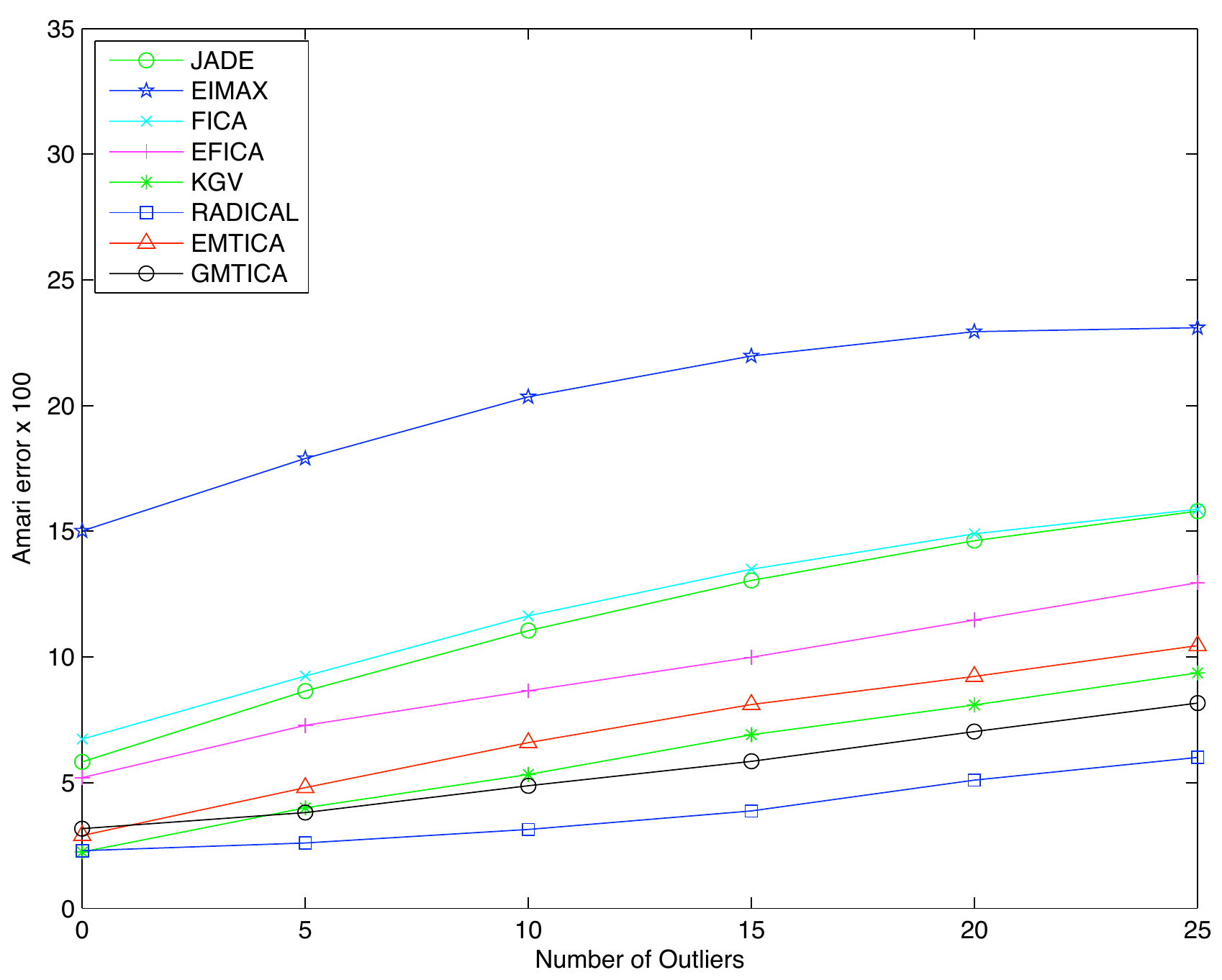,scale=0.7}}
\caption{Robustness to outliers. The averaged Amari errors obtained by the compared algorithms versus number of outliers for five-component ICA with randomly chosen source distributions. One sees that Gaussian-MTICA is less sensitive to outliers than exponential-MTICA. The RADICAL algorithm exhibits least sensitivity to outliers. However, this comes at the expense of increased computational complexity as indicated by Table \ref{ACLTAB} and the run time analysis in Table \ref{RunTimeTab}. }
\label{Fig6}
\end{figure}

In this experiment we demonstrate the robustness of the compared algorithms to outliers. We simulated outliers by randomly corrupting up to 25 data points out of the 1000 samples. This was carried out by adding the value $+5$ or $-5$, chosen with probability 1/2, to a single component in each of the selected data points. We performed 3000 Monte-Carlo simulations using source distributions chosen uniformly at random from the 10 possible distributions in Table \ref{DistributionsTable}. The averaged Amari errors produced by each algorithm are depicted in Fig. \ref{Fig6}. One can observe that, as expected, the proposed Gaussian-MTICA method is less sensitive to outliers than the exponential-MTICA. This is due to the boundedness of the Gaussian MT-function allowing it to de-emphasize outlying samples distant from its location parameter. The RADICAL algorithm exhibits the least sensitivity to outliers. However, this comes at the expense of significantly increased computational complexity as indicated by Table \ref{ACLTAB} and the run time analysis in Table \ref{RunTimeTab}.
\subsection{Sensitivity to model mismatch}
\label{ModelMismatch}
Here we demonstrate that exponential MT-ICA is more robust to model mismatch. To generate model mismatch we used the following noisy linear mixing model: 
\begin{equation}
\label{NoisyICAModel}
\Xmat=\Amat\Smat+\lambda\Emat,
\end{equation}
where $\Emat$ is a uniformly distributed additive noise vector with statistically independent components having zero mean and unit variance, and $\lambda>0$ is a scaling parameter that controls the signal-to-noise-ratio (SNR) according to 
\begin{equation} 
{\rm{SNR}}=\frac{{\rm{tr}\left[\Amat\Amat^{T}\right]}}{p\cdot\lambda^{2}}.
\end{equation}
For each value of SNR ranging from 0 [dB] to 12 [dB] we performed 3000 Monte-Carlo simulations. 

\begin{figure}[htbp!]
\centerline{\psfig{figure=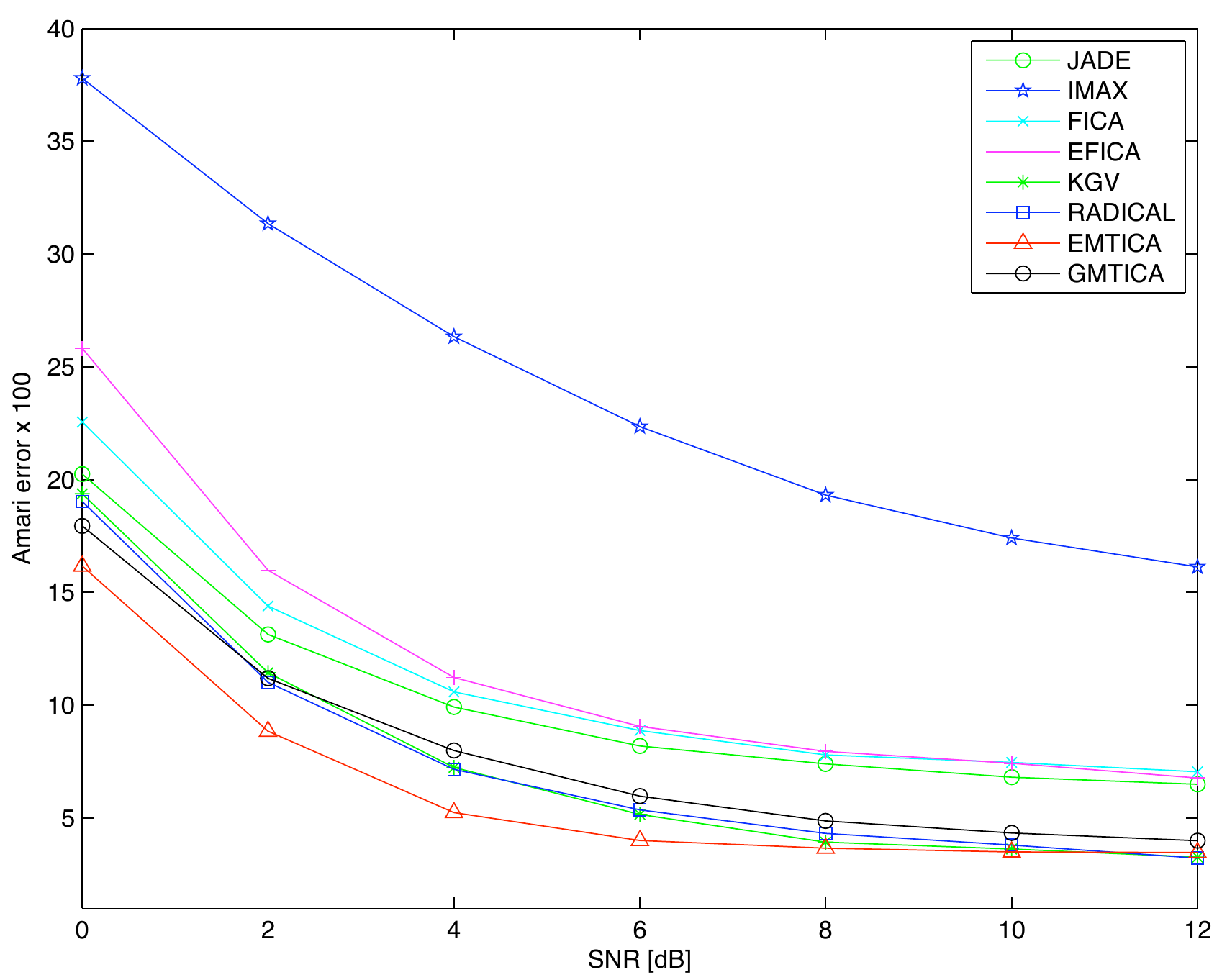,scale=0.7}}
\caption{Sensitivity to model mismatch. The averaged Amari errors obtained by the compared algorithms, under the noisy linear mixing model $\Xmat=\Amat\Smat+\lambda\Emat$, versus SNR. Since for low SNRs the whitened observations largely deviate from unitary mixing, exponential-MTICA outperforms all other algorithms that are based on whitening and unitary de-mixing. For high SNRs the whitened observations admit nearly unitary mixing, and thus, Gaussian-MTICA, KGV and RADICAL attain similar performance as compared to the exponential-MTICA.}
\label{Fig7}
\end{figure}

The averaged Amari errors obtained by each algorithm are depicted in Fig. \ref{Fig7}. Observe that for SNRs lower than 8 [dB] exponential-MTICA, which does not require whitening, outperforms all other compared algorithms that are based on whitening and unitary de-mixing. This is due to the fact that for low SNRs the whitened observations significantly deviates from unitary mixing. For higher SNRs one can notice that there is less of a separation performance gap between exponential-MTICA, Gaussian-MTICA, KGV and RADICAL. This is because for higher SNRs the whitened observations admit nearly unitary mixing. 
\section{Conclusion}
\label{Disc}
In this paper, a new framework for ICA was proposed that is based on applying a structured transform to the probability distribution of the data. In MTICA the separation matrix is estimated via approximate joint diagonalization of some empirical measure-transformed covariance matrices that are obtained by evaluating the MT-function at different test-points in the parameter space. By specifying the MT-function in the exponential family the ICA technique proposed in \citep{Yeredor}, called here exponential-MTICA, is obtained. Specification of the MT-function in the Gaussian family resulted in a new ICA algorithm called Gaussian-MTICA. The proposed MTICA approach was tested in simulation examples that illustrated the advantages of exponential-MTICA and Gaussian-MTICA over state-of-the-art algorithms for ICA. It is likely that there exist other classes of MT-functions that may result in other ICA algorithms using the proposed framework. 
\newpage
\appendix
\section{Proof Proposition \ref{Prop1}:}
\label{Prop1Proof}
\begin{enumerate}[\upshape(1\upshape)]
\item
\textbf{Property \ref{P1}:}\\
Since $\varphi_{u}\left(\xvec\right)$ is nonnegative, then by Corollary 2.3.6 in \citep{MeasureTheory} $\qx$ is a measure on $\mathcal{S}_{\XCalsc}$. Furthermore, $\qx\left(\XCal\right)=1$ so that $\qx$ is a probability measure on $\mathcal{S}_{\XCalsc}$. 
\item
\textbf{Property \ref{P2}:}\\
Follows from definitions 4.1.1 and 4.1.3 in \citep{MeasureTheory}.
\item
\textbf{Property \ref{P21}:}\\
According to the definition of $\varphi_{u}\left(\xvec\right)$ in (\ref{VarPhiDef}), the strict positivity of $u\left(\xvec\right)$, and Property \ref{P2}, we have that $\qx$ is absolutely continuous w.r.t. $\px$ with strictly positive Radon-Nikodym derivative $\frac{d\qx\left(\xvec\right)}{d\px\left(\xvec\right)}=\varphi_{u}\left(\xvec\right)$. Therefore, by Proposition 4.1.2 in \citep{MeasureTheory} it is implied that $\px$ is absolutely continuous w.r.t. $\qx$ with a strictly positive Radon-Nikodym derivative given by 
\begin{equation}
\label{EqP21}
\frac{d\px\left(\xvec\right)}{d\qx\left(\xvec\right)}=\varphi^{-1}_{u}\left(\xvec\right).
\end{equation}
Using (\ref{VarPhiDef}) and (\ref{EqP21}) one can easily verify that $\varphi^{-1}_{u}\left(\xvec\right)=\frac{u^{-1}\left(\xvec\right)}{{\rm{E}}\left[u^{-1}\left(\Xmat\right);\qx\right]}$.
\item
\textbf{Property \ref{P3}}:\\
Let $\qxk$ denote the marginal probability measure of $\qx$, defined on $\SCal_{\XCalsc_{k}}$. Additionally, let $A_{1},\ldots{A}_{p}$ denote arbitrary sets in the $\sigma$-algebras $\SCal_{\XCalsc_{1}},\ldots,\SCal_{\XCalsc_{p}}$, respectively. Using (\ref{Assumption1}), (\ref{MeasureTransform}), (\ref{VarPhiDef}), the assumed statistical independence of $X_{1},\ldots,X_{p}$ under $\px$, and Tonelli's Theorem \citep{Folland}:
\begin{eqnarray}
\label{Th1PrEq1}
\qx\left(A_{1}\times\cdots\times{A_{p}}\right)
&=&\int\limits_{A_{1}\times\cdots\times{A}_{p}}\frac{u\left(\xvec\right)}{{\rm{E}}\left[u\left(\Xmat\right);\px\right]}d\px\left(\xvec\right)
\\\nonumber
&=&\prod\limits_{k=1}^{p}\int\limits_{A_{k}}
\frac{u_{k}\left(x_{k}\right)}{{\rm{E}}\left[u_{k}\left(X_{k}\right);P_{X_{k}}\right]}dP_{X_{k}}\left(x_{k}\right),
\end{eqnarray}
which implies that 
\begin{equation}
\label{Th1PrEq2}
Q^{\left(u\right)}_{X_{k}}\left(A_{k}\right)=\qx\left(A_{k}\times\prod_{i\neq{k}}^{p}\XCal_{i}\right)=\int\limits_{A_{k}}\frac{u_{k}\left(x_{k}\right)}{{\rm{E}}\left[u_{k}\left(X_{k}\right);P_{X_{k}}\right]}dP_{X_{k}}\left(x_{k}\right).
\end{equation}
By (\ref{Th1PrEq1}) and (\ref{Th1PrEq2})
\begin{equation}
\label{Th1PrEq3}
\qx\left(A_{1}\times\cdots\times{A_{p}}\right)=\prod\limits_{k=1}^{p}\qxk\left(A_{k}\right).
\end{equation}
Therefore, since $A_{1},\ldots{A}_{p}$ are arbitrary, $X_{1},\ldots,X_{p}$ are mutually independent under the transformed probability measure $\qx$.
\end{enumerate}
\section{Proof of Proposition \ref{RobustnessConditions}:}
\label{RobustnessConditionsProof}
The influence function (\ref{MT_COV_INF}) can be written as:
\begin{equation}
{\rm{IF}}_{\textrm{H}_{u},\px}\left(\yvec\right)=\frac{1}{{\rm{E}}\left[u\left(\Xmat\right);\px\right]}
\left(\left(\sqrt{u\left(\yvec\right)}{\|\yvec\|}_{2}+\sqrt{u\left(\yvec\right)}{\|\muvec^{\left(u\right)}_{\Xmatsc}\|}_{2}\right)^{2}\psivec\left(\yvec\right)
\psivec^{T}\left(\yvec\right) - u\left(\yvec\right){\bSigma}^{\left(u\right)}_{\Xmatsc} \right),
\end{equation}
where
\begin{equation}
\psivec\left(\yvec\right)\triangleq\frac{\yvec-\muvec^{\left(u\right)}_{\Xmatsc}}{{\|\yvec\|}_{2}+{\|\muvec^{\left(u\right)}_{\Xmatsc}\|}_{2}}.
\end{equation}
By the triangle inequality ${\left\|\psivec\left(\yvec\right)\right\|}_{2}\leq{1}$ for any $\yvec\in\Rsp^{p}$, and therefore, the matrix term $\psivec\left(\yvec\right)
\psivec^{T}\left(\yvec\right)$ is bounded. Thus, the influence function ${\rm{IF}}_{\textrm{H}_{u},\px}\left(\yvec\right)$ is bounded if $u\left(\yvec\right)$ and $u\left(\yvec\right)\left\|\yvec\right\|^{2}_{2}$ are bounded. \qed
\section{Proof of Theorem \ref{ConstVarExpMT}:}
\label{ExpMTVarConstProof}
Define ${M}^{\left(h_{\rm{E}}\right)}_{X}\left(t\right)\triangleq{\rm{E}}\left[\exp\left(tX\right);Q^{\left(h_{\rm{E}}\right)}_{X}\right]$
as the moment generating function of $X$ under the transformed probability measure $Q^{\left(h_{\rm{E}}\right)}_{X}$ that is associated with the exponential MT-function 
\begin{equation}
\label{ExpMTFunc2}
h_{\rm{E}}\left(X;t_{0}\right)=\exp\left(t_{0}X\right).
\end{equation}
Using (\ref{VarPhiDef}), (\ref{MTCovZ}), (\ref{ExpMTFunc}) and (\ref{ExpMTFunc2}) one can verify that
\begin{equation}
\label{VarMGFExpRelation}
\sigma^{\left(\uexp\right)}_{X}\left(t+t_{0}\right)=\frac{\partial^{2}\log{M}^{\left(h_{\rm{E}}\right)}_{X}\left(t\right)}{\partial{t^{2}}}.
\end{equation}

If the condition in (\ref{ExpMTVarConst}) is satisfied then by (\ref{VarMGFExpRelation}) and the properties of the moment generating function
\begin{equation}
\label{MGFExp}
{M}^{\left(h_{\rm{E}}\right)}_{X}\left(t\right)=\exp\left(\mu^{\left(h_{\rm{E}}\right)}_{X}{t}+\frac{1}{2}\sigma^{\left(h_{\rm{E}}\right)}_{X}t^{2}\right)
\hspace{0.2cm}\forall{t}\in\left(-\epsilon,\epsilon\right),
\end{equation}
where $\mu^{\left(h_{\rm{E}}\right)}_{X}$ and $\sigma^{\left(h_{\rm{E}}\right)}_{X}$ denote the mean and variance of $X$ under $Q^{\left(h_{\rm{E}}\right)}_{X}$, respectively. Since the moment generating function, reduced to any open interval that contains the origin, uniquely determines the distribution \citep{Severini, DasGupta}, then $Q^{\left(h_{\rm{E}}\right)}_{X}$ is a Gaussian probability measure. Hence, by Lemma \ref{ExpPresProp} in Subsection \ref{ExpMTCov} we conclude that $P_{X}$ is Gaussian.

Conversely, if $P_{X}$ is Gaussian then by Lemma \ref{ExpPresProp} in Subsection \ref{ExpMTCov} the probability measure $Q^{\left(h_{\rm{E}}\right)}_{X}$ is Gaussian, and its corresponding moment generating function ${M}^{\left(h_{\rm{E}}\right)}_{X}\left(t\right)$ must satisfy (\ref{MGFExp}). Therefore, using (\ref{VarMGFExpRelation}) one obtains 
$\sigma^{\left(\uexp\right)}_{X}\left(t\right)=\sigma^{\left(h_{\rm{E}}\right)}_{X}$ $\forall{t}\in\left(t_{0}-\epsilon,t_{0}+\epsilon\right)$. \qed
\section{Proof of Theorem \ref{ZeroLebesgueExpTh}:}
\label{ZeroLebesgueExpThProof}
We need to prove that 
$$\left\{\left(\tvec_{1},\tvec_{2}\right)\in\Rsp^{p}\times\Rsp^{p}:\tvec_{1}\neq\tvec_{2}\hspace{0.1cm}{\rm{and}}
\hspace{0.1cm}\Lambdamat^{\left(u_{\rm{E}}\right)}_{\Smatsc}\left(\Amat^{T}\tvec_{1},\Amat^{T}\tvec_{2}\right){\hspace{0.1cm}\rm{does\hspace{0.1cm}not\hspace{0.1cm}
have\hspace{0.1cm}distinct\hspace{0.1cm}diagonal\hspace{0.1cm}entries}}\right\}$$ 
has zero Lebesgue measure. 
Since $\muvec=\Amat^{T}\tvec$ defines a bijective mapping from $\Rsp^{p}$ to $\Rsp^{p}$ this is equivalent to showing that the Lebesgue measure of the set
\begin{equation}
\nonumber
\mathcal{D}\triangleq\left\{\left(\muvec_{1},\muvec_{2}\right)\in\Rsp^{p}\times\Rsp^{p}
:\muvec_{1}\neq\muvec_{2}\hspace{0.1cm}{\rm{and}}
\hspace{0.1cm}
\Lambdamat^{\left(u_{\rm{E}}\right)}_{\Smatsc}\left(\muvec_{1},\muvec_{2}\right)\hspace{0.1cm}{\rm{does\hspace{0.1cm}not\hspace{0.1cm}have\hspace{0.1cm}distinct\hspace{0.1cm}diagonal\hspace{0.1cm}entries}}\right\}
\end{equation}
is zero. By the definition of $\Lambdamat^{\left(u_{\rm{E}}\right)}_{\Smatsc}\left(\muvec_{1},\muvec_{2}\right)$ in Proposition \ref{ExpAIdent}, the set $\mathcal{D}$ can be written as 
\begin{equation}
\label{DExpDecomp}
\mathcal{D}=\bigcup\limits_{{j}\neq{k}}^{p}\mathcal{D}_{j,k},
\end{equation}
where
\begin{equation}
\label{DjkExp}
\mathcal{D}_{j,k}\triangleq\left\{\left(\muvec_{1},\muvec_{2}\right)\in\Rsp^{p}\times\Rsp^{p}:\muvec_{1}\neq\muvec_{2}\hspace{0.2cm}{\rm{and}}\hspace{0.2cm}
\frac{\sigma^{\left(\uexp\right)}_{S_{j}}\left(\mu_{1,j}\right)}{\sigma^{\left(\uexp\right)}_{S_{j}}\left(\mu_{2,j}\right)}=
\frac{\sigma^{\left(\uexp\right)}_{S_{k}}\left(\mu_{1,k}\right)}{\sigma^{\left(\uexp\right)}_{S_{k}}\left(\mu_{2,k}\right)}\right\},
\end{equation}
$\sigma^{\left(\uexp\right)}_{S_{j}}\left(\mu_{i,j}\right)=\left[\bSigma^{\left(\uexp\right)}_{\Smatsc}\left(\muvec_{i}\right)\right]_{j,j}$, and $\mu_{i,j}=\left[\muvec_{i}\right]_{j}$, $i=1,2$, $j=1,\ldots,p$. Since at most one of the sources is Gaussian, 
then either $S_{j}$ or $S_{k}$ must be non-Gaussian for $j\neq{k}$. Let $S_{k}$ denote the non-Gaussian source. By Theorem \ref{ConstVarExpMT}, the exponential MT-variance $\sigma^{\left(\uexp\right)}_{S_{k}}\left(\mu_{1,k}\right)$ is not constant over any open interval. Thus, for almost every $\left(\mu_{1,j},\mu_{2,j},\mu_{1,k},\mu_{2,k}\right)\in\Rsp^{4}$ for which the quotients in (\ref{DjkExp}) are finite $\frac{\sigma^{\left(\uexp\right)}_{S_{j}}\left(\mu_{1,j}\right)}{\sigma^{\left(\uexp\right)}_{S_{j}}\left(\mu_{2,j}\right)}\neq\frac{\sigma^{\left(\uexp\right)}_{S_{k}}\left(\mu_{1,k}\right)}{\sigma^{\left(\uexp\right)}_{S_{k}}\left(\mu_{2,k}\right)}$.
Hence, the Lebesgue measure of $\mathcal{D}_{j,k}$ is zero for any $j\neq{k}$. Therefore, by relation (\ref{DExpDecomp}) and the sub-additivity of Lebesgue's measure, the set $\mathcal{D}$ must have zero Lebesgue measure.
\qed
\section{Proof of Theorem \ref{ConstVarGaussMT}:}
\label{ConstVarGaussMTProof}
Define ${M}^{\left(h_{\rm{G}}\right)}_{X}\left(t\right)\triangleq{\rm{E}}\left[\exp\left(tX\right);Q^{\left(h_{\rm{G}}\right)}_{X}\right]$ 
as the moment generating function of $X$ under the transformed probability measure $Q^{\left(h_{\rm{G}}\right)}_{X}$ associated with the Gaussian MT-function 
\begin{equation}
\label{GaussMTFunc2}
h_{\rm{G}}\left(X;t_{0},\tau\right)=\exp\left(-\frac{\left(X-t_{0}\right)^{2}}{2\tau^{2}}\right).
\end{equation}
Using (\ref{VarPhiDef}), (\ref{MTCovZ}), (\ref{GaussKernel}) and (\ref{GaussMTFunc2}) one can verify that
\begin{equation}
\label{VarMGFRelation}
\sigma^{\left(\uGausss\right)}_{X}\left(t+t_{0},\tau\right)=\tau^{4}\frac{\partial^{2}\log{M}^{\left(h_{\rm{G}}\right)}_{X}\left(t/\tau^{2}\right)}{\partial{t^{2}}}.
\end{equation}

If the condition in (\ref{GaussMTVarConst}) is satisfied then by (\ref{VarMGFRelation}) and the properties of the moment generating function
\begin{equation}
\label{MGFGauss}
{M}^{\left(h_{\rm{G}}\right)}_{X}\left(t\right)=\exp\left(\mu^{\left(h_{\rm{G}}\right)}_{X}{t}+\frac{1}{2}\sigma^{\left(h_{\rm{G}}\right)}_{X}t^{2}\right)
\hspace{0.2cm}\forall{t}\in\left(-\epsilon^{\prime},\epsilon^{\prime}\right),
\end{equation}
where $\mu^{\left(h_{\rm{G}}\right)}_{X}$ and $\sigma^{\left(h_{\rm{G}}\right)}_{X}$ denote the mean and the variance of $X$ under $Q^{\left(h_{\rm{G}}\right)}_{X}$, respectively, and $\epsilon^{\prime}\triangleq\epsilon/\tau^{2}$. Since the moment generating function, reduced to any open interval that contains the origin, uniquely determines the distribution \citep{Severini, DasGupta} it is implied that $Q^{\left(h_{\rm{G}}\right)}_{X}$ is a Gaussian probability measure. Hence, by Lemma \ref{GaussPresProp} in Subsection \ref{GaussMTCov} we conclude that $P_{X}$ is Gaussian.

Conversely, if $P_{X}$ is Gaussian then by Lemma \ref{GaussPresProp} in Subsection \ref{GaussMTCov} the probability measure $Q^{\left(h_{\rm{G}}\right)}_{X}$ is Gaussian, and its corresponding moment generating function ${M}^{\left(h_{\rm{G}}\right)}_{X}\left(t\right)$ must satisfy (\ref{MGFGauss}). Therefore, using (\ref{VarMGFRelation}) one obtains 
$\sigma^{\left(\uGausss\right)}_{X}\left(t,\tau\right)=\sigma^{\left(h_{\rm{G}}\right)}_{X}$ $\forall{t}\in\left(t_{0}-\epsilon,t_{0}+\epsilon\right)$. \qed
\section{Proof of Theorem \ref{ZeroLebesgueGauss}:}
\label{ZeroLebesgueGaussProof}
We need to prove that 
$\left\{\tvec\in\Rsp^{p}:\bSigma^{\left(\uGausss\right)}_{\Smatsc}\left(\Umat^{T}\tvec,\tau\right)\hspace{0.1cm}{\rm{does\hspace{0.1cm}not\hspace{0.1cm}have\hspace{0.1cm}distinct\hspace{0.1cm}diagonal\hspace{0.1cm}entries}}\right\}$ 
has zero Lebesgue measure. Since the relation $\muvec=\Umat^{T}\tvec$ defines a bijective mapping from $\Rsp^{p}$ to $\Rsp^{p}$ it is sufficient to show that the Lebesgue measure of the set 
\begin{equation}
\mathcal{D}\triangleq\left\{\muvec\in\Rsp^{p}:\bSigma^{\left(\uGausss\right)}_{\Smatsc}\left(\muvec,\tau\right)\hspace{0.1cm}{\rm{does\hspace{0.1cm}not\hspace{0.1cm}
have\hspace{0.1cm}distinct\hspace{0.1cm}diagonal\hspace{0.1cm}entries}}\right\}
\end{equation}
is zero. Clearly, the set $\mathcal{D}$ can be written as 
\begin{equation}
\label{DGaussDecomp}
\mathcal{D}=\bigcup\limits_{{j}\neq{k}}^{p}\mathcal{D}_{j,k},
\end{equation}
where
\begin{equation}
\mathcal{D}_{j,k}\triangleq\left\{\muvec\in\Rsp^{p}:\sigma^{\left(\uGausss\right)}_{S_{j}}\left(\mu_{j},\tau\right)=
\sigma^{\left(\uGausss\right)}_{S_{k}}\left(\mu_{k},\tau\right)\right\},
\end{equation}
$\sigma^{\left(\uGausss\right)}_{S_{j}}\left(\mu_{j}\right)=\left[\bSigma^{\left(\uGausss\right)}_{\Smatsc}\left(\muvec,\tau\right)\right]_{j,j}$ and $\mu_{j}=\left[\muvec\right]_{j}$.  Since at most one of the sources is Gaussian, then either $S_{j}$ or $S_{k}$ must be non-Gaussian for $j\neq{k}$. Let $S_{k}$ denote the non-Gaussian source. By Theorem \ref{ConstVarGaussMT} the Gaussian MT-variance  $\sigma^{\left(\uGausss\right)}_{S_{k}}\left(\mu_{k},\tau\right)$ is not constant w.r.t. $\mu_{k}$ over any open interval. Thus, for almost every $\left(\mu_{j},\mu_{k}\right)\in\Rsp^{2}$ we have that $\sigma^{\left(\uGausss\right)}_{S_{j}}\left(\mu_{j},\tau\right)\neq\sigma^{\left(\uGausss\right)}_{S_{k}}\left(\mu_{k},\tau\right)$. Hence, the Lebesgue measure of $\mathcal{D}_{j,k}$ is zero for any $j\neq{k}$. Therefore, by relation (\ref{DGaussDecomp}) and the sub-additivity of Lebesgue's measure, the set $\mathcal{D}$ must have zero Lebesgue measure.
\qed
\section{Choice of MT-function parameters}
\subsection{\textbf{Exponential MTICA}}
\label{EMTICA_PARAM} 
Assume that $\tvec_{1},\ldots,\tvec_{M}$ are independent samples from some continuous probability distribution. According to Theorem \ref{ZeroLebesgueExpTh} if at most one of the sources is Gaussian, then for any pair $\left(\tvec_{m},\tvec_{n}\right)$, $m\neq{n}$, Assumption \ref{A3_ExpAIdent} in Proposition \ref{ExpAIdent} is satisfied with probability 1 that leads to unique identification of $\Amat$ based on the corresponding MT-covariance matrices $\bSigma^{\left(\uexp\right)}_{\Xmatsc}\left(\tvec_{m}\right)$ and $\bSigma^{\left(\uexp\right)}_{\Xmatsc}\left(\tvec_{n}\right)$. 

Motivated by this result we propose the following procedure that randomly generates test-points inside a unit $l_{2}$-ball: 
\begin{enumerate}[\upshape(1\upshape)]
\item
Generate $M$ i.i.d samples $\rvec_{m}\in\Rsp^{p}$, $m=1,\ldots,M$ from the standard normal distribution.
\item
Generate $M$ i.i.d. samples $c_{m}\in\Rsp$, $m=1,\ldots,M$ with uniform distribution on $\left[0,1\right]$.
\item
Obtain the sequence of test-points:
$$\tvec_{m}=c_{m}\frac{\rvec_{m}}{\left\|\rvec_{m}\right\|_{2}},\hspace{0.1cm}m=1,\ldots,M.
$$
\end{enumerate}
\subsection{\textbf{Gaussian MTICA}}
\label{GMTICA_PARAM}
Assume that $\tvec_{1},\ldots,\tvec_{M}$ are independent samples from some continuous probability distribution. According to Theorem \ref{ZeroLebesgueGauss} if at most one of the sources is Gaussian, then for any $m=1,\ldots,M$ the Gaussian MT-covariance $\bSigma^{\left(\uGausss\right)}_{\Zmatsc}\left(\tvec_{m},\tau\right)$ in (\ref{MTCovGauss}) has distinct eigenvalues with probability 1 that leads to unique identification of the mixing matrix $\Amat$. 

Motivated by this result, and assuming that the data is centered and whitened, we propose to generate $M$ i.i.d. vectors $\tvec_{m}$, $m=1,\ldots,M$, such that the components of each $\tvec_{m}$ are statistically independent with zero mean and unit variance. In all considered examples we used the beta distribution with identical shape parameters $\alpha=\beta=3$.
\bibliography{mybib}
\end{document}